\documentclass[aps,twocolumn,showpacs,preprintnumbers,amsmath,amssymb,superscriptaddress,prr,longbibliography]{revtex4-1}

\usepackage{graphicx}
\usepackage{dcolumn}
\usepackage{bm}
\usepackage{graphics}
\usepackage{subfigure}
\usepackage{graphicx}
\usepackage{epsfig}
\usepackage{epstopdf}
\usepackage{xcolor}
\usepackage{multirow}
\usepackage{mathtools}

\allowdisplaybreaks

\begin{document}

\title{Collective modes in interacting two-dimensional tomographic Fermi liquids}

\author{Johannes Hofmann}
\email{johannes.hofmann@physics.gu.se}
\affiliation{Department of Physics, Gothenburg University, 41296 Gothenburg, Sweden}

\author{Sankar Das Sarma}
\email{dassarma@umd.edu}
\affiliation{Department of Physics, University of Maryland, 41296 College Park, USA}

\date{\today}

\begin{abstract}
We develop an analytically solvable model for interacting two-dimensional Fermi liquids with separate collisional relaxation rates for parity-odd and parity-even Fermi surface deformations. Such a disparity of collisional lifetimes exists whenever scattering is restricted to inversion-symmetric Fermi surfaces, and should thus be a generic feature of two-dimensional Fermi liquids. It implies an additional unanticipated ``tomographic'' transport regime (in between the standard collisionless and hydrodynamic regimes) in which even-parity modes are overdamped while odd-parity modes are collisionless. We derive expressions for both the longitudinal and the transverse conductivity and discuss the collective mode spectrum along the collisionless-tomographic-hydrodynamic crossover. Longitudinal modes cross over from zero sound in the collisionless regime to hydrodynamic first sound in the tomographic and hydrodynamic regime, where odd-parity damping appears as a subleading correction to the lifetime. In charged Fermi liquids with long-range Coulomb coupling, these modes reduce to plasmons with a strongly suppressed odd-parity correction to the damping. The transverse response, by contrast, has a specific tomographic transport regime with two imaginary odd-parity modes, one of which requires a finite repulsive interaction, distinct from both the shear sound in the collisionless regime and an overdamped diffusive current mode in the hydrodynamic limit. Our work demonstrates that there are deep many-body aspects of interacting Fermi liquids, which are often thought to be well understood theoretically, remaining unexplored.
\end{abstract}

\maketitle

\section{Introduction}

Interaction-dominated hydrodynamic electron transport has recently become accessible in low-dimensional Fermi liquids~\cite{bandurin16,crossno16,moll16,nam17,krishnakumar17,gooth18,gusev18,bandurin18,braem18,berdyugin19,gallagher19,sulpizio19,jenkins22} (with important earlier work~\cite{dejong95,buhmann02}). Observing such a hydrodynamic transport regime requires both strong electron interactions as well as clean samples in which the rate of impurity scattering $\gamma_i$ is much smaller than the electron collision rate $\gamma$, \mbox{$\gamma_i \ll \gamma$}, i.e., electron-electron interactions thermalize the system. Standard phase-space arguments predict \mbox{$\gamma \sim (T/T_F)^2$}~\cite{baym04,giuliani05,pines18} (for \mbox{$T\ll T_F$}, where $T$ is the temperature and $T_F$ is the Fermi temperature, valid up to logarithmic corrections~\cite{giuliani82,zheng96,li13,dassarma21}), such that electron interactions dominate at higher temperatures until other mechanisms (such as phonon scattering) become relevant; in graphene, for example, the hydrodynamic regime exists around \mbox{$T=100$-$150$K}~\cite{bandurin16}, and in bilayer graphene around \mbox{$T= 50$-$100$K}~\cite{bandurin16,bandurin18}. Depending on the ratio between the quasiparticle scattering rate $\gamma$ and the characteristic transport frequency (set by $v q$, where $v$ is the Fermi velocity and $q$ the wave number of an excitation), the interaction-dominated Fermi liquid is either in a collisionless regime, \mbox{$\gamma \ll v q$}, where particle scattering is not efficient to damp excitations over the time scale of the mode, or in a hydrodynamic regime, \mbox{$\gamma\gg vq$}, where only zero modes of the collision integral participate in the dynamics while all other modes are overdamped. Although such a collision-dominated hydrodynamic regime is not of much relevance to normal metals, by virtue of \mbox{$T_F\sim10^4$K} in regular metals, hydrodynamics should be the generic behavior of many clean 2D materials, such as monolayer~\cite{bandurin16} and bilayer graphene~\cite{bandurin18},  and high-mobility 2D GaAs structures~\cite{ahn22}.

\begin{figure}[b!]
\scalebox{1.2}{\includegraphics{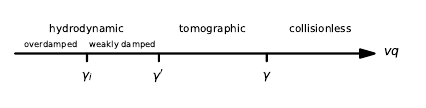}}\\[-5ex]
\caption{Sketch of the distinct transport regimes in a two-dimensional Fermi liquid with weak disorder. Here, $\gamma$, $\gamma'$, and $\gamma_i$ are, respectively, the even-parity, the odd-parity, and the impurity relaxation rate.}
\label{fig:1}
\end{figure}

However, this canonical picture of Fermi liquids described above is actually not complete for two-dimensional materials. Here, the restriction of quasiparticle excitations to the vicinity of a two-dimensional Fermi surface imposes stronger constraints compared to the three-dimensional case~\cite{laikhtman92,gurzhi95,ledwith17,ledwith19,ledwith19b,hofmann22}. In particular, for Fermi surfaces with inversion symmetry \mbox{${\bf p} \leftrightarrow - {\bf p}$}, the predicted quadratic temperature dependence of the collision rate only applies to the parity-even part of the quasiparticle distribution $f_+({\bf p}) = [f({\bf p}) + f(-{\bf p})]/2$~\cite{gurzhi95,nilsson05}. The parity-odd part $f_-({\bf p}) = [f({\bf p}) - f(-{\bf p})]/2$, by contrast, is seen to decay much more slowly, which defines an additional collisional scale $\gamma' \ll \gamma$. Indeed, for a circular Fermi surface (which we consider in this paper), one finds $\gamma' \sim (T/T_F)^4$~\cite{ledwith19,hofmann22}, a result that on a microscopic level is interpreted in terms of angle-correlated two-body scattering on the Fermi surface~\cite{ledwith17}. The emergence of two separate interaction-induced collisional lifetimes raises the interesting prospect of a completely new ``tomographic'' transport regime, in which parity-even deformations of the Fermi surface are overdamped yet all parity-odd modes are still collisionless; see Fig.~\ref{fig:1} for an illustration of the various limits. The system is both collision-dominated (``even-parity'') and collision-free (``odd-parity'') in the tomographic regime. Predicted observables include a scale-dependent shear viscosity~\cite{ledwith19b}, with possible signatures reflecting a deviation from the laminar hydrodynamic Poiseuille flow profile~\cite{ledwith19b,sulpizio19}, or enhanced backscattering around impurities~\cite{hong20}.

In this work, we provide a comprehensive discussion of collective excitations in a two-dimensional Fermi liquid along the frequency-tuned collisionless-tomographic-hydrodynamic crossover. To this end, we formulate an analytically solvable model for the quasiparticle dynamics that includes  impurity scattering as well as the two collisional timescales through a relaxation-time ansatz. The model is solvable since at low temperatures changes in the quasiparticle distribution function---\mbox{$\delta f_{{\bf p}}^{}(t, {\bf r}) = f_{\bf p}(t,{\bf r}) - f_0(\varepsilon_{\bf p})$}, where ${\bf p}$ is the wave number of an excitation with single-particle energy $\varepsilon_{{\bf p}}$ and $f_0$ the equilibrium Fermi-Dirac function---is strongly peaked at the Fermi surface ($|{\bf p}| = p_F$ for the circular Fermi surfaces that we consider) such that the dynamics reduces to the angular dynamics of the Fermi surface deformation,
\begin{align}
\delta f^{}(\theta; t, {\bf r}) &= \biggl(- \frac{\partial f_0}{\partial \varepsilon_{{\bf p}}}\biggr) \delta \mu(\theta; t, {\bf r}) ,
\end{align}
where $\delta \mu$ is a time- and position-dependent variation in the chemical potential and $\theta$ is the angle that parametrizes the position of the momentum on the Fermi surface. If expanded in angular harmonics,
\begin{align}
\delta \mu(\theta) &= \sum_m \delta \mu_m e^{im\theta} , \label{eq:deformation}
\end{align}
the continuous kinetic Fermi liquid equation reduces to a discrete tight-binding form, where even (odd) $m$ describe even-parity (odd-parity) deformations of the Fermi surface, with \mbox{$m=0$} describing density fluctuations and \mbox{$m=\pm1$} current fluctuations. We consider a simple relaxation-time ansatz in the angular components to account for the two parity-dependent collisional relaxation times,
\begin{align}
\gamma_m &= \begin{cases}
0 & m=0 , \\
\gamma_i & |m|=1 , \\
\gamma & m \ {\rm even}, \, |m|\geq2 , \\
\gamma' & m \ {\rm odd}, \, |m|\geq3 ,
\end{cases} \label{eq:damping}
\end{align}
where we impose $\gamma_i < \gamma' < \gamma$ as shown in Fig.~\ref{fig:1} as appropriate for a clean interaction-dominated system. In addition, to describe short-range Fermi liquid interactions, we include the dominant Landau parameters in our calculations, which are the isotropic parameter $F_0$ for the longitudinal response (which sets the compressibility) and the dipole parameter $F_1$ for the transverse response (which sets the mass renormalization). This is of course a simplified description, but it allows an analytic discussion of the collective mode spectrum and it contains all salient physics. The model~\eqref{eq:damping} is accurate for semiconductors at low temperature or semimetals at finite doping such as graphene~\cite{hofmann22} (with a breakdown of the odd-even effect near charge neutrality~\cite{kiselev19,kiselev20}). Indeed, it is straightforward to extend our model to include more interaction parameters or more complicated relaxation terms (as is discussed in the Appendix), and we demonstrate that the structure of the modes remains unchanged. We note that our formalism follows the classic prescription for describing the interacting system using a few phenomenological parameters following the procedure originally outlined by Landau in his famous Landau Fermi liquid theory~\cite{landau57a,landau57b}.

Before presenting a full derivation, we summarize in Sec.~\ref{sec:results} the main results for the collective mode spectrum for longitudinal and transverse excitations both in charged and neutral Fermi liquids, and discuss the behavior along the hydrodynamic-tomographic-collisionless crossover. Our main finding for the longitudinal collective mode is a direct crossover from collisionless zero sound to hydrodynamic first sound with subleading signatures of parity-dependent damping, which is replaced by a collective plasmon mode with similar damping for charged Fermi liquids. For the transverse collective mode, three distinct transport regimes exist, with a collisionless shear sound mode, two imaginary tomographic modes, and a diffusive hydrodynamic current mode. Section~\ref{sec:model} presents a detailed discussion of the kinetic description of the Fermi liquid, derives the longitudinal and transverse current response function, the poles of which determine the collective modes, and discusses the causal structure of the response. 
The paper concludes in Sec.~\ref{sec:summary}.

\section{Results}\label{sec:results}

\begin{figure}[t!]
\scalebox{1}{\includegraphics{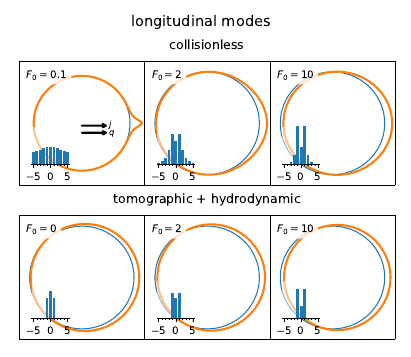}}
\caption{
Fermi surface deformation for longitudinal excitations in the collisionless regime (top row) and the tomographic and hydrodynamic regime (bottom row) with increasing strength of the Fermi liquid parameter $F_0$ (left to right). The insets show the decomposition of the mode in angular components. In the collisionless regime, at small parameter values the mode is strongly localized around the direction of propagation, while at large parameters, the mode reduces to a longitudinal dipole (\mbox{$m=\pm1$}) mode. In the tomographic and hydrodynamic regime, the mode predominantly involves the density and current zero modes.
}
\label{fig:2}
\end{figure}

We begin by discussing longitudinal collective modes, in which the current is parallel to the direction of the wave number. For a collisionless neutral Fermi liquid, where \mbox{$\gamma_i < \gamma' < \gamma < v q$} (cf. Fig.~\ref{fig:1}), we find a weakly damped collective zero sound mode that splits from the particle-hole continuum  for $F_0\geq 0$:
\begin{align}
\omega &= \pm \frac{1+F_0}{\sqrt{1+2F_0}} v q - i \frac{1}{(1+2F_0)^2} \biggl(\frac{1+2F_0}{2} \gamma \nonumber \\
&\qquad + \frac{1}{2} \gamma' + 2 F_0 (1+F_0)\gamma_i\biggr) + {\it O}\biggr(\frac{1}{q}\biggr) . \label{eq:Lcoll}
\end{align}
The real part is the standard zero sound mode of a two-dimensional Fermi liquid~\cite{anderson11,khoo19,klein19,torre19} with an imaginary part that sets the damping of the mode. Figure~\ref{fig:2} (top row) shows the corresponding Fermi surface deformation $\delta \mu(\theta)$ for several values of $F_0$, where the insets show the spectral decomposition as a function of the angular parameter~$m$. At small values of $F_0$, the zero sound mode involves all angular components and is sharply localized around the direction of propagation. For large $F_0$, by contrast,  the mode reduces to a pure dipole (\mbox{$m=\pm1$}) oscillation. This structure is reflected in the damping of the mode: For small $F_0$, the leading-order damping is set by the even-parity relaxation $\gamma$ with a subleading correction $\gamma' $. For large $F_0$, however, the impurity scattering is dominant as ${\it O}(1)$ while the even-mode relaxation term $\gamma$ is suppressed as ${\it O}(1/F_0)$ (due to the small \mbox{$m=\pm2$} component) and the odd-mode relaxation as ${\it O}(1/F_0^2)$ (due to the even smaller \mbox{$m=\pm3$} component).
{\it O}
The longitudinal response does not have a distinct tomographic signature. Both in the tomographic regime, \mbox{$\gamma_i < \gamma' < v q < \gamma$}, and in the hydrodynamic regime, \mbox{$\gamma_i < v q < \gamma' < \gamma$}, the mode reduces to a damped hydrodynamic first sound mode
\begin{align}
\omega &= \pm \sqrt{\frac{1+F_0}{2}} v q - i \biggl(\frac{\gamma_i}{2} + \frac{(vq)^2}{8 \gamma}\biggr) + {\it O}\biggl(\frac{(vq)^3}{\gamma^2}\biggr) , \label{eq:longhydro}
\end{align}
which exists for all \mbox{$F_0 \geq -1$} above the Pomeranchuk instability. The interaction contribution to the damping is of the standard hydrodynamic form $-i \nu q^2/2$, where $\nu = v^2/4\gamma$ is the kinetic expression for the viscosity of a 2D Fermi liquid~\cite{guo16,lucas18}. For an illustration of the collective mode structure, see the bottom plot of Fig.~\ref{fig:2}, which only involves the zero-mode components \mbox{$m=0$} and \mbox{$m=\pm1$}. Finally, in the overdamped hydrodynamic limit \mbox{$v q < \gamma_i < \gamma' < \gamma$}, the mode reduces to a density diffusion mode with
\begin{align}
\omega &= - i \frac{1+F_0}{2} \frac{(vq)^2}{\gamma_i} ,
\end{align}
which is again not sensitive to the odd-parity damping~$\gamma'$.

\begin{figure}[t!]
\scalebox{1}{\includegraphics{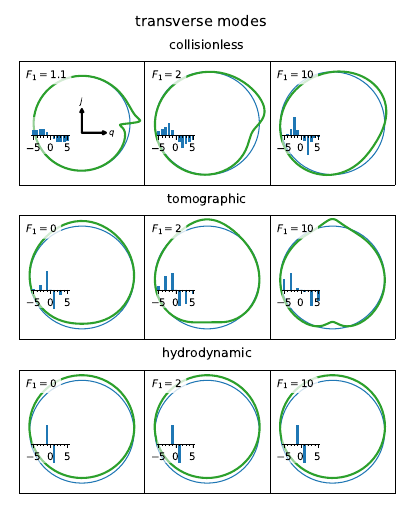}}
\caption{
Fermi surface deformation for transverse excitations in the collisionless (top row), the tomographic (middle row), and the hydrodynamic regime (bottom row) with increasing strength of the Fermi liquid parameter $F_1$ (left to right). The insets show the decomposition of the mode in angular components. 
In the collisionless regime, at small parameter values the mode is strongly localized around the direction of momentum and transverse to the direction of propagation, while at large parameters, the mode reduces to a transverse quadrupole oscillation. In the tomographic limit, even-parity modes are overdamped with no contribution of the density mode, but odd modes remain. Their contribution increases at large repulsion while the contribution of the current zero mode is reduced. In the hydrodynamic regime, only the current zero mode contributes. 
}
\label{fig:3}
\end{figure}

The above discussion applies to charge-neutral systems with short-range bare interactions, which only exist in normal He-$3$~\cite{abel66} and atomic quantum gases~\cite{froehlich12,yan19}. For charged systems, Landau-Silin theory includes an internal polarization field in the Fermi-liquid description that accounts for the dynamical screening due to the long-range Coulomb interaction. This is essential for the convergence of the theory in the presence of long-range interactions. For a plane-wave perturbation with wave number~$q$, the net effect of the Coulomb interaction is a shift in the isotropic Landau parameter $F_0$~\cite{lucas18},
\begin{align}
F_0 \to F_0 + \frac{2 \pi \alpha}{\lambda_F q} , \label{eq:coulombshift}
\end{align}
where $\lambda_F$ is the Fermi wavelength and $\alpha = e^2/\kappa \hbar v$ is the dimensionless Coulomb interaction strength (i.e., the effective ``fine structure constant'') with $\kappa$ a dielectric constant. Equation~\eqref{eq:coulombshift} can be thought of as the screening introduced by the long-range part of the Coulomb interaction, with the long-wavelength divergence for \mbox{$q\to0$}, in going from the Landau theory to the Landau-Silin theory. Results for neutral Fermi liquids can thus be applied directly to charged systems, for which only the limit $F_0\to \infty$ is relevant (since $q \ll 1/\lambda_F$ within the validity of the Fermi liquid theory). Both the zero sound mode in the collisionless regime and the first sound mode in the hydrodynamic regime then become a collective plasmon mode~\cite{lucas18}. In the collisionless regime, we have
\begin{align}
\omega &= \pm \sqrt{\frac{\pi \alpha v^2}{\lambda_F}} \sqrt{q} - i \biggl(\frac{\gamma_i}{2} + \frac{\gamma}{8 \pi \alpha} \lambda_F q + \frac{\gamma'}{32 \pi^2 \alpha^2} (\lambda_F q)^2\biggr) .
\end{align}
The real part is the standard plasmon frequency of a two-dimensional electron gas~\cite{ando82,hwang07,dassarma09}, with a damping that receives an odd-parity anomalous contribution, which is distinguished not only by a distinct temperature scaling (through the anomalous temperature dependence of $\gamma'$) but also by a characteristic ${\it O}(q^2)$ momentum dependence. The hydrodynamic plasmon is~\cite{lucas18}
\begin{align}
\omega &= \pm \sqrt{\frac{\pi \alpha v^2}{\lambda_F}} \sqrt{q} - i \biggl(\frac{\gamma_i}{2} + \frac{(vq)^2}{8 \gamma}\biggr) ,
\end{align}
with a damping that is unchanged from Eq.~\eqref{eq:longhydro} and that has no odd-parity correction at this order.

Let us now discuss the transverse collective mode, for which the current flow is perpendicular to the momentum direction. Unlike the longitudinal response, the transverse modes do not involve a density oscillation and are thus not sensitive to the isotropic Landau parameter~$F_0$. Our results thus apply equally to both neutral and charged Fermi liquids since screening is not a relevant consideration for the  transverse modes. The transverse response has two distinct imaginary modes in the tomographic regime that separate the regime from the collisionless limit, which has a collective shear sound mode for sufficiently strong repulsion, and the hydrodynamic limit, which has a diffusive current mode. 
 
 First, the transverse collisionless shear sound mode, which exists for $F_1\geq 1$, is
\begin{align}
\omega &= \pm \frac{1+F_1}{2 \sqrt{F_1}} v q - i \biggl(\frac{\gamma}{2} + \frac{\gamma'}{2 F_1} + \frac{F_1 - 1}{2 F_1} \gamma_i\biggr) + {\it O}\biggr(\frac{1}{q}\biggr) . \label{eq:transversefree}
\end{align}
The speed of sound is not affected by collisions~\cite{khoo19,conti99,gao10}. The structure of the mode is illustrated in the top row of Fig.~\ref{fig:3}. Unlike longitudinal excitations, transverse modes are antisymmetric and do not involve a density (\mbox{$m=0$}) component because the direction of current flow is perpendicular to the momentum direction.  For small Fermi liquid parameters $F_1$, the mode is strongly peaked in the momentum direction with a large number of higher harmonics. With increasing Landau parameter it reduces to a pure quadrupole (\mbox{$m=\pm2$}) oscillation. The even-parity damping in Eq.~\eqref{eq:transversefree} is thus not suppressed compared to the impurity damping at large $F_1$. The odd-parity contribution, however, is suppressed at large $F_1$ as before. 

In the tomographic regime, the shear sound mode is replaced by two purely imaginary low-energy modes. For $\gamma_i=0$, their dispersion is
\begin{align}
\omega &= - i \gamma' \frac{1+F_1}{8 F_1} \biggl( 4 + (q\xi)^2 (1+F_1) \nonumber \\
&\qquad\quad \mp (q\xi)^2 \sqrt{\biggl(1+F_1 + \frac{4}{(q\xi)^2}\biggr)^2 - \frac{16 F_1}{(q\xi)^2}}\biggr) , \label{eq:diffusivetomo}
\end{align}
where we define a characteristic length scale
\begin{align}
\xi = \frac{v}{\sqrt{\gamma\gamma'}} . \label{eq:xi}
\end{align}
Note that the dimensionless quantity $q\xi$ can take any positive value in the tomographic regime, where large values approach the collisionless limit and small values the hydrodynamic limit, cf. Fig.~\ref{fig:1}.  An impurity scattering rate $\gamma_i$ is included in Eq.~\eqref{eq:diffusivetomo} by shifting the overall dispersion by $-i\gamma_i$ and replacing $\gamma' \to \gamma' - \gamma_i$. The minus sign in Eq.~\eqref{eq:diffusivetomo} defines an upper branch with frequency \mbox{$-i \gamma' < \omega < 0$} that exists for all \mbox{$-1 < F_1$}, and the plus sign denotes a lower branch with frequency $\omega < -i \gamma' (1+(q\xi)^2)$ that starts at \mbox{$F_1^* = 1+ 2/(q\xi)^2$}. The two modes are separated by a branch cut on the imaginary axis that extends from $-i\gamma'(1+(q\xi)^2)$ to $-i\gamma'$.

\begin{figure}[t!]
\scalebox{1.2}{\includegraphics{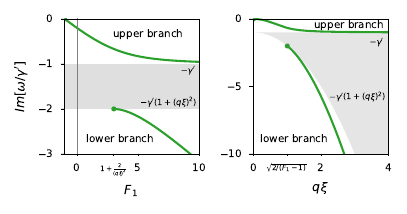}}\vspace{-0.5cm}
\caption{
Damping of the two tomographic transverse modes for \mbox{$\gamma_i = 0$} (a) as a function of the Fermi liquid parameter $F_1$ for fixed $q\xi = 1$ and (b) as a function of $q\xi$ for fixed \mbox{$F_1=3$}. The gray shaded area indicates the branch cut of the transverse response.
}
\label{fig:4}
\end{figure}

The two tomographic modes in Eq.~\eqref{eq:diffusivetomo} are sketched in Fig.~\ref{fig:4}(a) for \mbox{$q\xi=1$} and $\gamma_i=0$ as a function of $F_1$. The upper-branch dispersion starts out linearly near $F_1 \simeq -1$ as 
\begin{align}
\omega &= - i \gamma' \frac{1+F_1}{2} \Bigl(\sqrt{1+(q\xi)^2} - 1\Bigr) ,
\end{align}
takes at $F_1 = 0$ the value
\begin{align}
\omega &= - i \gamma' \frac{(q\xi)^2}{4+(q\xi)^2} ,
\label{eq:upper0}
\end{align}
and behaves asymptotically for $F_1 \to \infty$ as 
\begin{align}
\omega &= - i \gamma' . \label{eq:upperinf}
\end{align}
The latter limit is independent of the impurity relaxation rate $\gamma_i$. The lower branch starts at the branch point \mbox{$-i\gamma'(1+(q\xi)^2)$} and diverges for large $F_1$ as
 \begin{align}
 \omega &= - i \frac{F_1}{4} \frac{(vq)^2}{\gamma} , \label{eq:lowerinf}
 \end{align}
 thus clearly separating from the upper mode, Eq.~\eqref{eq:upperinf}. Seen as a function of $q\xi$ (Fig.~\ref{fig:4}(b)), the upper branch starts out at small $q\xi$ with a quadratic dispersion \mbox{$\omega = - i (1+F_1) (vq)^2/4\gamma$} characteristic of a diffusive mode, but is then superdiffusive for larger $q\xi$ (i.e., with effective exponent \mbox{$\alpha < 2$}) and levels off to \mbox{$\omega = - i \gamma'$} at large $q\xi$. The lower mode approaches the branch point for smaller $q\xi$ (i.e., when approaching the hydrodynamic limit) and vanishes at a critical value $q^*\xi = \sqrt{2/(F_1-1)}$, such that only the upper branch remains.
 
 The Fermi surface deformation of the upper branch is shown in the middle row of Fig.~\ref{fig:3} for \mbox{$q\xi=1$} and several values of $F_1$. For small $F_1$, the mode is predominantly a dipole mode with small weight in the odd-parity harmonics. Increasing the Landau parameter $F_1$ (or the momentum $q\xi$) increases the weight of the odd-parity components such that the damping approaches the  value~\eqref{eq:upperinf}. By contrast, the lower-branch diffusion mode (not shown) is dominated by the zero-modes \mbox{$m=0,\pm 1$} at large $F_1$ or large $q\xi$ with hydrodynamic damping~\eqref{eq:lowerinf}.  As becomes clear from Figs.~\ref{fig:3} and~\ref{fig:4}, there are two competing damping mechanisms in the tomographic regime: the direct damping $-i\gamma'$ of the odd-parity modes and the hydrodynamic damping $-i(vq)^2/4\gamma$. 

Finally, in the hydrodynamic limit, the upper-branch mode in Eq.~\eqref{eq:diffusivetomo} reduces to a diffusive current mode
\begin{align}
\omega = - i \biggl(\gamma_i + \frac{1+F_1}{4} \frac{(vq)^2}{\gamma} \biggl) + {\it O}\biggl(\frac{(vq)^3}{\gamma^2}\biggr) ,
\end{align}
with a direct damping due to impurity scattering and a hydrodynamic damping term $\sim \nu q^2$. For $\gamma_i=0$ and \mbox{$F_1=0$}, this result agrees with~\cite{guo16}. The mode structure for different $F_1$ is shown in the bottom row of Fig.~\ref{fig:3}, which only involves the current components and hence does not change with $F_1$. Just as the longitudinal hydrodynamic first sound mode, the transverse hydrodynamic mode is not sensitive to the odd-parity damping at this order.

\section{Fermi liquid model}\label{sec:model}

In this section, we review the kinetic Fermi-liquid description of the angular dynamics (Sec.~\ref{sec:IIIa}) and derive analytical expressions for the longitudinal and transverse optical conductivity (Sec.~\ref{sec:IIIb}), the poles of which determine the collective modes discussed in the previous section. In addition, Sec.~\ref{sec:IIIc} provides further details on the analytic structure of the response functions.

\subsection{Kinetic equation and tight-binding form}\label{sec:IIIa}

The time evolution of $\delta f^{}(\theta; t, {\bf r})$ is governed by the quasiparticle kinetic equation~\cite{baym04,pines18}
\begin{align}
\frac{\partial \delta f(\theta)}{\partial t} - \frac{\partial \tilde{\varepsilon}}{\partial {\bf r}} \cdot \frac{\partial \delta f(\theta)}{\partial {\bf p}} + \frac{\partial \tilde{\varepsilon}}{\partial {\bf p}} \cdot \frac{\partial \delta f(\theta)}{\partial {\bf r}} &= {\cal J}[\delta f(\theta)] , \label{eq:kineticabstract}
\end{align}
where the left-hand side describes free phase space evolution based on the local quasiparticle energy
\begin{align}
\tilde{\varepsilon}_{\theta}(t, {\bf r}) &= \varepsilon_{{\bf p}} + \int_0^{2\pi} \frac{d\theta'}{2\pi} F(\theta - \theta') \delta \mu(\theta'; t, {\bf r}) , \label{eq:localqpenergy}
\end{align}
which includes a mean field shift due to (short-range) interactions parametrized by the Landau function $F(\theta)$. The right-hand side ${\cal J}[\delta f_\theta]$ of Eq.~\eqref{eq:kineticabstract} is the collision integral, which captures gains and losses due to scattering between particles or particles and impurities. Substituting Eq.~\eqref{eq:localqpenergy} in Eq.~\eqref{eq:kineticabstract} gives
\begin{align}
\frac{\partial}{\partial t} \delta f(\theta) + {\bf v}_\theta \cdot \frac{\partial}{\partial {\bf r}} \delta \bar{f}(\theta) + \frac{\partial f_0}{\partial {\bf p}} \cdot (-e {\bf E}) = {\cal J}[\delta f(\theta)] , \label{eq:kinetic}
\end{align}
where ${\bf v}_{\theta} = \partial \varepsilon_{\bf p}/\partial {\bf p}|_{p_F}$ is the Fermi velocity, $\delta \bar{f}(\theta) = f(\theta) - f_0(\tilde{\varepsilon}_\theta)$ is the deviation from local thermodynamic equilibrium, which is determined by the local quasiparticle energy~\eqref{eq:localqpenergy}, and we include an external field ${\bf E}$ that generates a driving force. Explicitly, in terms of the chemical potential variation $\delta \bar{f}(\theta) = (- \partial f_0/\partial \varepsilon_{{\bf p}}) \delta \bar{\mu}(\theta)$,
\begin{align}
\delta \bar{\mu}(\theta) &= \delta \mu(\theta) + \int_0^{2\pi} \frac{d\theta'}{2\pi} F(\theta - \theta') \delta \mu(\theta') .
\end{align}
For a plane-wave perturbation 
\begin{align}
{\bf E}_{\rm ext}(t, {\bf r}) &= {\bf E}_0 \, e^{- i \omega t + i {\bf q} \cdot {\bf r}}
\end{align}
with corresponding Fermi surface response $\delta \mu(\theta; t, {\bf r}) = e^{- i \omega t + i {\bf q} \cdot {\bf r}} \delta \mu(\theta)$, the kinetic equation reduces to (${\bf q}$ is aligned along the $x$ axis)
\begin{align}
&\bigl( - \omega + v q \cos \theta\bigl) \delta \mu(\theta) - i v E_0 \cos (\theta-\theta_E) \nonumber \\
&\quad+ v q \cos \theta \, 
\int_0^{2\pi} \frac{d\theta'}{2\pi} \, \tilde{F}(\theta-\theta') \, \delta \mu(\theta')  = - i {\cal L}[\delta \mu] , \label{eq:linearizedkinetic}
\end{align}
where ${\cal L}[\delta \mu] = {\cal J}[\delta f(\theta)]/(- \partial f_0/\partial \varepsilon_{{\bf p}})$ is the linearized collision integral and $\theta_E$ the angle between the external field and the $x$ axis.

The kinetic equation~\eqref{eq:linearizedkinetic} is converted to a tight-binding form by expanding in angular harmonics as in Eq.~\eqref{eq:deformation}, expanding the Landau function in the same way as
\begin{align}
F(\theta) &= \sum_m F_m e^{im\theta} ,
\end{align}
as well as replacing the linearized collision integral by its smallest eigenvalue,
\begin{align}
{\cal L}[\delta \mu_m] &= - \gamma_m \delta \mu_m .
\end{align}
This gives
\begin{align}
&(- \omega - i \gamma_{m}) \delta \mu_m + \frac{v q}{2} \bigl(1 + F_{m-1}\bigr) \delta \mu_{m-1} \nonumber \\
&\qquad+ \frac{v q}{2} \bigl(1 + F_{m+1}\bigr) \delta \mu_{m+1} \nonumber \\
&=  - \frac{i v E_0}{2} \bigl(e^{-i \theta_E} \delta_{m,+1} + e^{i \theta_E} \delta_{m,-1}\bigr) . \label{eq:kinetictightbinding}
\end{align}
Formally, this equation takes the form of a one-dimensional tight-binding Hamiltonian with lattice sites $m$ and wave function $\delta \mu_m$, where the term $(- \omega - i \gamma_{m})$ is a complex on-site energy and the $v q (1 + F_{m\pm1})/2$ terms are tunneling amplitudes between nearest neighbors from site \mbox{$m\pm1$} to $m$. A single nonzero Fermi liquid parameter modifies the hopping amplitude between two sites and is interpreted as a bond impurity. Increasing $F_0$, for example, increases the hopping from site $0$ to sites $\pm1$, while increasing $F_1$ increases the hopping from site $\pm 1$ to the sites $0$ and $\pm 2$. This structure is immediately apparent in the solutions shown in Figs.~\ref{fig:2} and~\ref{fig:3}. The general numerical solution of the tight-binding equation~\eqref{eq:kinetictightbinding} is discussed in the Appendix.

\subsection{Response functions}\label{sec:IIIb}

The current ${\bf j}$ induced in response to an applied electric field is linked to the deviation from local thermodynamic equilibrium~\cite{baym04,pines18},
\begin{align}
{\bf j} &= \int \frac{d{\bf p}}{(2\pi)^2} \, {\bf v}_{{\bf p}} \delta \bar{f}_{{\bf p}}^{}(t, {\bf r}) \nonumber \\
&= \nu_0 \frac{v}{2} (1+F_1)
\begin{pmatrix}
\delta \mu_{+1} + \delta \mu_{-1} \\[1ex]
i (\delta \mu_{+1} - \delta \mu_{-1})
\end{pmatrix} ,
\end{align}
where $\nu_0=m^*/\pi$ is the 2D density of states and $m^*$ the effective mass. Note that Galilean invariance implies \mbox{$m^*/m = 1+F_1$}. 
We extract $\delta \mu_{\pm1}$ from the kinetic equation~\eqref{eq:kinetictightbinding} restricted to the hydrodynamic modes $m=0,\pm1$,
\begin{align}
&\begin{pmatrix}
- \omega - i \gamma_i + \frac{v q}{2} x_2 & \frac{v q}{2} (1 + F_0) & 0 \\[1ex]
\frac{v q}{2} (1 + F_1) & - \omega & \frac{v q}{2} (1 + F_1) \\[1ex]
0 & \frac{v q}{2} (1 + F_0) & - \omega - i \gamma_i + \frac{v q}{2} x_{2}
\end{pmatrix} \nonumber \\[1ex]
&\qquad \times \begin{pmatrix}
\delta \mu_{1} \\[1ex]
\delta \mu_0 \\[1ex]
\delta \mu_{-1}
\end{pmatrix}
=
- \frac{v}{2} 
\begin{pmatrix}
e^{i \theta_E} \\[1ex] 
0 \\[1ex] 
e^{- i \theta_E}
\end{pmatrix} E_0 ,
\end{align}
where $x_2 = \delta \mu_2/\delta \mu_1$ incorporates the coupling to higher harmonics. As discussed below, $x_2$ follows from the kinetic equation~\eqref{eq:kinetictightbinding} without reference to the source terms, where $x_2=0$ defines the hydrodynamic limit. In terms of this quantity, the longitudinal and transverse conductivity read
\begin{align}
\frac{\sigma_\parallel(\omega, {\bf q})}{\nu_0} &= - \frac{q^2 v^2 (1+F_1)}{2} \, \nonumber \\
&\qquad  \times\frac{1}{- \omega (\omega+i\gamma_i) + c_1^2 q^2 + \frac{v q}{2} \omega x_2} \label{eq:sigmaL} \\
\frac{\sigma_\perp(\omega, {\bf q})}{\nu_0} &= + \frac{\omega^2 v^2 (1+F_1)}{2} \, \frac{1}{- \omega (\omega+i\gamma_i) + \frac{v q}{2} \omega x_2} , \label{eq:sigmaT}
\end{align}
where we define the speed of first sound
\begin{align}
\frac{c_1^2}{v^2} &= \frac{(1+F_0) (1+F_1)}{2} .
\end{align}
To obtain the full analytic response functions, it is sufficient to compute the coefficient $x_2$ with the damping~\eqref{eq:damping}.
 In general form, the kinetic equation~\eqref{eq:kinetictightbinding} for $m\geq 2$ yields a continued fraction representation of $x_2$. This representation of the tight-binding problem can be solved for any damping $\gamma_m$ with only minor numerical cost as discussed in the Appendix. For the odd-even damping given in Eq.~\eqref{eq:damping} that we discuss in this work, it is even possible to obtain an analytical result: In this case, $x_2$ is written in the form
\begin{align}
x_2 &= - \dfrac{\tfrac{vq}{2} (1+F_1)}{- \omega - i \gamma + \dfrac{- (\tfrac{vq}{2})^2}{- \omega - i \gamma' + \dfrac{- (\tfrac{vq}{2})^2}{- \omega - i \gamma+\ldots}}} \nonumber \\
&= - \dfrac{\tfrac{vq}{2} (1+F_1)}{- \omega - i \gamma + A} ,
\end{align}
where $A$ fulfills
\begin{align}
A &= \dfrac{- (\tfrac{vq}{2})^2}{- \omega - i \gamma' + \dfrac{- (\tfrac{vq}{2})^2}{- \omega - i \gamma + A}} .
\end{align}
The solution is
\begin{align}
A &= \frac{1}{2} \biggl(\omega + i \gamma - \sqrt{(\omega + i \gamma)^2 - v^2 q^2 \, \frac{\omega + i \gamma}{\omega + i \gamma'}}\biggr) ,
\end{align}
which in turn yields
\begin{align}
x_2 &= \frac{v q (1+F_1)}{\omega + i \gamma + \sqrt{(\omega + i \gamma)^2 - v^2 q^2 \, \frac{\omega + i \gamma}{\omega + i \gamma'}}} . \label{eq:x2full}
\end{align}
The poles of the response functions~\eqref{eq:sigmaL} and~\eqref{eq:sigmaT} evaluated with the expression~\eqref{eq:x2full} yield the collective modes discussed in Sec.~\ref{sec:results}. In deriving these results, it is helpful to use dimensionless scaling variables $s=\omega/vq$, $\Gamma=\gamma/vq$, $\Gamma'=\gamma'/vq$, and $\Gamma_i=\gamma_i/vq$.

\begin{figure}[t!]
\scalebox{1}{\includegraphics{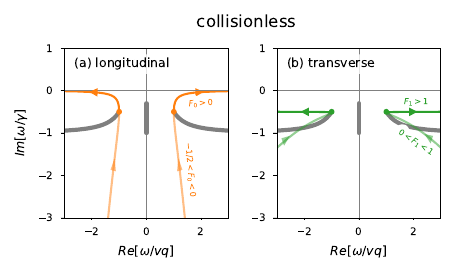}}
\caption{
Analytic structure of the (a) longitudinal and (b) transverse response in the collisionless limit \mbox{$\gamma_i \ll \gamma' \ll \gamma \ll vq$}. Thick gray lines indicate the branch cuts, and orange and green lines show the zero sound and shear sound poles as a function of the Landau parameters $F_0$ and $F_1$, respectively. For attractive interactions \mbox{$-1<F_0<-1/2$} and \mbox{$-1<F_1<0$}, the modes are purely imaginary and lie on the physical and unphysical sheet. For \mbox{$-1/2<F_0<0$} and \mbox{$0<F_1<1$}, sound modes exists but they are hidden behind the branch cut and do not give rise to a resonance in the response function (transparent lines). At $F_0=0$ and $F_1=1$, the modes cross the branch points and become the zero sound and shear sound modes discussed in Sec.~\ref{sec:results} (full lines).
}
\label{fig:5}
\end{figure}

\subsection{Analytic structure}\label{sec:IIIc} 

In addition to the main results for the collective mode spectrum presented in Sec.~\ref{sec:results}, we include in this section a discussion of the analytic structure of the response functions~\eqref{eq:sigmaL} and~\eqref{eq:sigmaT} in the complex frequency plane. Through Eq.~\eqref{eq:x2full}, the response contains a square-root function of the frequency. This function has a two-sheeted Riemann surface with a physical sheet on which the square-root function has positive real part and an unphysical sheet with negative real part~\cite{klein20}. Poles on the physical sheet are located in the lower half of the complex plane to ensure causality.

In the collisionless regime, the response function has two branch points at $\pm vq - i (\gamma + \gamma')/2$ with branch cuts that extends to infinity. An additional branch cut extends from $-i\gamma'$ to $-i\gamma$ on the imaginary axis. This structure is shown in Fig.~\ref{fig:5}, where Fig.~\ref{fig:5}(a) shows the longitudinal collective modes for all $F_0$ as orange lines and Fig.~\ref{fig:5}(b) shows the transverse collective modes for all $F_1$ as green lines, where arrows indicate the direction of increasing Landau parameter. Full lines indicate the zero sound and shear sound modes above the branch cut stated in Eqs.~\eqref{eq:Lcoll} and~\eqref{eq:transversefree}, respectively. These modes show up as resonances in the response and are separated from the particle-hole continuum. For smaller values of the interaction (\mbox{$-1/2<F_0<0$} and \mbox{$0<F_1<1$}, respectively), the modes are strongly overdamped sound modes that lie behind the branch cut. Such modes were dubbed ``hidden modes'' in Ref.~\cite{klein20}. Note that even though the mode for \mbox{$0<F_1<1$} in Fig.~\ref{fig:5}(b) crosses the branch cut, it lies on the unphysical sheet and is thus still hidden. They will not show up as a resonance (i.e.,~``overdamped'') in the response but they will impact the time evolution of perturbations and thus could be observed in pump-probe experiments. For even smaller values above the Pomeranchuk instability \mbox{$F_{0/1} > -1$}, the modes lie on the imaginary axis, both in the negative complex plane on the physical sheet and on the positive complex plane on the unphysical sheet.

\begin{figure}[t!]
\scalebox{1}{\includegraphics{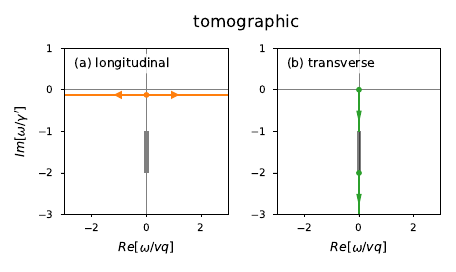}}
\scalebox{0.85}{\includegraphics{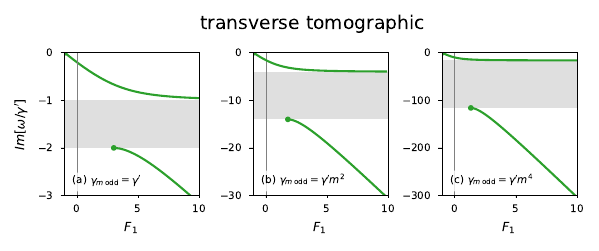}}
\caption{
Top row: Complex structure of the (a) longitudinal and (b) transverse response in the tomographic limit \mbox{$\gamma_i \ll \gamma' \ll vq \ll \gamma$}. Thick gray lines indicate the branch cut, and orange and green lines show the first sound and diffusive poles as a function of the Landau parameters $F_0$ and $F_1$, respectively. The longitudinal first sound mode has a fixed damping independent of the Landau parameter, and the transverse response has two diffusive modes with an upper and a lower branch separated by a branch cut that extends between $(\gamma', \gamma'+(vq)^2/\gamma)$. Bottom row: Frequency of the diffusive transverse modes as a function of $F_1$ for three different odd-parity damping mechanisms: (a) $\gamma_{m, {\rm odd}} = \gamma'$ (as discussed in Sec.~\ref{sec:results}), (b) $\gamma_{m, {\rm odd}} = \gamma' m^2$, and (c) $\gamma_{m, {\rm odd}} = \gamma' m^4$, with similar results.
}
\label{fig:6}
\end{figure}

The branch cut structure depends on the wave number of the excitation and thus changes for the different transport regimes. In the tomographic regime, the only branch cut that remains at low energies lies on the imaginary axis and extends from $-i\gamma'$ to $-i\gamma'(1+(q\xi)^2)$ as discussed after Eq.~\eqref{eq:xi} (the other branch cut remains of order ${\it O}(-i\gamma)$). The corresponding trajectory of tomographic modes is shown in the top row of Fig.~\ref{fig:6}, where the orange lines indicate the first sound mode~\eqref{eq:longhydro} and the green lines the upper and lower transverse tomographic modes~\eqref{eq:diffusivetomo}.

A natural question is whether this analytic structure survives more complicated damping terms. For example, Ref.~\cite{ledwith19b} proposes an $m$-dependent odd-parity relaxation rate of the form
\begin{align}
\gamma_{m,\,{\rm odd}} &= \gamma' m^p ,
\end{align}
valid at least for small $m$, which reflects an effective angular relaxation dynamics of the odd-parity distribution~\cite{ledwith17}. For diffusive dynamics \mbox{$p=2$}, and for subdiffusive dynamics \mbox{$p=4$}. For $p\neq0$, an analytic solution for the collective mode spectrum does not exist, but a numerical solution as outlined in the Appendix is straightforward. Our calculations confirm that the structure of the response and the tomographic modes do not change: We find a branch cut in the response that extends on the negative imaginary axis between $-ip^2\gamma'$ and $-i\gamma'(p^2+(q\xi)^2)$. An upper branch exists for all $-1<F_1$ with the same structure as discussed in Sec.~\ref{sec:results}, leveling off to $-ip^2\gamma'$. The lower branch decouples from the lower branch point at $F_1^* = 1 + 2 p^2/(q\xi)^2$. This structure is illustrated in the bottom row of Fig.~\ref{fig:6}, which shows the tomographic modes for three damping models (a) $p=0$ (discussed in Sec.~\ref{sec:results}), (b) $p=2$, and (c) $p=4$. This illustrates that the excitation spectrum is accurately predicted by the analytically solvable model discussed in this paper.

\section{Summary}\label{sec:summary}

We have discussed an analytically solvable model for an interaction-dominated two-dimensional Fermi liquid that includes a parity-dependent collisional relaxation rate, which is solved in closed form since the quasiparticle dynamics at low temperatures reduces to the angular dynamics of the Fermi surface deformation. In formulating the model, we have of course simplified the theoretical description and neglected, for example, additional Landau parameters or more complicated relaxation rates. However, as demonstrated in this paper, it is straightforward, if necessary, to include such modifications in a numerical solution, which only yields quantitative corrections. Additional corrections to the model discussed here arise at higher temperatures for which the energy dependence of the Fermi surface deformation $\delta \mu$ can no longer be absorbed in an effective relaxation-time ansatz for the~$\gamma_m$. Extending the present analysis using a systematic basis expansion that includes this energy dependence of Fermi surface perturbations~\cite{hofmann22} is an interesting prospect for further study.

Let us briefly comment on the possible observability of our predictions. For neutral Fermi liquids, longitudinal sound and sound damping have been measured in He-$3$~\cite{abel66} as well as atomic Fermi gases~\cite{vogt12,patel20}. Likewise, the plasmon dispersion and their damping in charged two-dimensional systems has been observed using a variety of techniques~\cite{allen77,pinczuk81,pinczuk86,bostwick07,liu08,tegenkamp10,bostwick10,ju11}. 
However, the predicted signatures arising from two separate parity-dependent collisional scales are subleading corrections to the damping and thus likely difficult to extract. By contrast, these signatures are more pronounced in the transverse response, for which we predict a separate tomographic transport regime. Shear sound in a collisionless 3D Fermi liquid has been measured in He-$3$~\cite{roach76}, although these results are debated~\cite{flowers76}. The excitation and detection of transverse modes remains a challenge as they do not involve a charge excitation (and hence do not easily couple to external probes), although recent proposals suggest that they strongly modify the conductivity in narrow channels~\cite{khoo20}, including a strong enhancement of transmission of terahertz radiation~\cite{valentinis21}. In the tomographic regime, the upper-branch mode is predicted to exist for all values of the Landau parameter and should thus be a generic feature of two-dimensional Fermi liquids that does not require any fine tuning. The lower-branch tomographic mode requires strong repulsion \mbox{$F_1 > 1+ 2/(q\xi)^2$}, corresponding to a many-body Fermi velocity renormalization of at least by a factor of $2$ in the short-wavelength limit. However, this criterion is no worse than that for the observability of standard shear sound~\cite{khoo19} and indeed much smaller than the critical interaction strength in three-dimensional systems~\cite{valentinis21}. In addition, our results indicate that the two competing time scales $\tau_1 = 1/\gamma'$ and $\tau_2 = \gamma/(vq)^2$ in the transverse tomographic response could be revealed in pump-probe experiments, which detect the real-time evolution of the current: The real-time current $j(t)$ following a quench is given by the Fourier transform of Eq.~\eqref{eq:sigmaT}, and the long-time behavior will thus be determined by the low-energy poles and branch cuts of the response~\cite{klein20}. The two time scales $\tau_1$ and $\tau_2$, which can be varied by changing both wave number and temperature, set the decay rate of the current at early and late times. \\

{\it Note added:} Recently, version 3 of Ref.~\cite{kryhin22} appeared, which discusses collective modes in a model restricted to \mbox{$m=1$} and \mbox{$m=3$} harmonics.

\begin{acknowledgments}
J.H. thanks Ulf Gran and F. Setiawan for discussions. This work is supported by Vetenskapsr\aa det (Grant No. 2020-04239). The work at the University of Maryland is supported by the Laboratory for Physical Sciences.
\end{acknowledgments}

\appendix

\section{Numerical solution}\label{app:A}

In this Appendix, we outline a general numerical solution method for the kinetic Fermi liquid equation without uncontrolled truncation errors. The method is valid for arbitrary damping and thus applicable even if an analytical solution cannot be found. The key is a continued fraction representation of the tight-binding form~\eqref{eq:kinetictightbinding} of the kinetic equation, for which efficient exact numerical methods exist. For further details and related applications, we refer to the discussion in Ref.~\cite{setiawan21}. 

Formally, the kinetic equation~\eqref{eq:kinetictightbinding} is a linear system of equations of the form
\begin{align}
&\gamma_{m}^{-} \delta \mu_{m-1} + \alpha_m \delta \mu_m + \gamma_{m}^{+} \delta \mu_{m+1} = \beta_1 \delta_{m,1} + \beta_{-1} \delta_{m,-1} , \label{eq:tightbinding}
\end{align}
where the coefficients are defined implicitly from Eq.~\eqref{eq:kinetictightbinding}.  The left-hand side is a sparse matrix with entries on the diagonal and the first upper and lower bands. In principle, one could truncate this matrix equation at a large cutoff index $\pm m_{\rm max}$ setting $\delta \mu_{\pm m_{\rm max}} = 0$ and iterate backward, which however introduces uncontrolled truncation errors and will certainly fail in the collisionless and tomographic transport regime in which there is a significant contribution of higher modes (cf. Figs.~\ref{fig:2} and~\ref{fig:3}). To construct the full solution, consider first the homogenous components for $|m| \geq 2$ and define
\begin{align}
x_m = \begin{cases}
\dfrac{\delta \mu_m}{\delta \mu_{m-1}} & m\geq2 \\[2ex]
\dfrac{\delta \mu_m}{\delta \mu_{m+1}} & m\leq-2 ,
\end{cases}
\end{align}
in terms of which Eq.~\eqref{eq:tightbinding} reads
\begin{alignat}{2}
\frac{\gamma_{m}^{-}}{x_{m}} + \alpha_m + \gamma_{m}^{+} x_{m+1} &= 0 \qquad &m&\geq 2 \\
\gamma_{m}^- x_{m-1} + \alpha_m + \frac{\gamma_{m}^+}{x_{m}} &= 0 \qquad &m&\leq -2 .
\end{alignat}
These two equations can be written as a continued fraction for a given $m$:
\begin{align}
x_m &= \begin{cases}
\dfrac{- \gamma_m^-}{\alpha_m + \gamma_m^+ x_{m+1}} & m\geq 2\\[3ex]
\dfrac{- \gamma_m^+}{\alpha_m + \gamma_m^- x_{m-1}} & m\leq -2 ,
\end{cases} \label{eq:continuedfraction}
\end{align}
or, in general form,
\begin{align}
x_m &= b_{0} + \dfrac{a_{1}}{b_{1} + \dfrac{a_{2}}{b_{2} + \dfrac{a_{3}}{b_{3}+\ldots}}} \label{eq:continuedfraction}
\end{align}
with coefficients (where $m\geq2$):
\begin{align}
b_{p} &= \begin{cases}
0 & p=0 \\[1ex]
\alpha_{m+(p-1)} & p>0
\end{cases} \\[1ex]
a_{p} &= \begin{cases}
- \gamma_m^- & p=1 \\[1ex]
- \gamma_{m+(p-1)}^+ \gamma_{m+p}^- & p>1
\end{cases}
\end{align}
and for $m\leq-2$:
\begin{align}
b_{p} &= \begin{cases}
0 & p=0 \\[1ex]
\alpha_{m-(p-1)} & p>0
\end{cases} \\[1ex]
a_{p} &= \begin{cases}
- \gamma_m^+ & p=1 \\[1ex]
- \gamma_{m-(p-1)}^- \gamma_{m-p}^+ & p>1
\end{cases} .
\end{align}
This continued fraction is solved exactly using the modified Lentz method~\cite{press17}, which we summarize in the following: Define by $x_m^{(n)}$ the approximate solution for $x_m$ obtained by truncating the partial fraction expansion for $x_m$ at some index $m+n$, $x_{n+m} = 0$. For a given $m$, this approximation may be expressed as a rational fraction (suppressing all $m$ indices in the following)
\begin{align}
x^{(n)} &= \frac{A_n}{B_n}
\end{align}
with
\begin{align}
A_n &= b_n A_{n-1} + a_n A_{n-2} \label{eq:An} \\
B_n &= b_n B_{n-1} + a_n B_{n-2} \label{eq:Bn}
\end{align}
and initial conditions $A_{-1} = 1, A_0 = b_{0}, B_{-1} = 0$ and $B_0 = 1$ (the proof of the above statement proceeds by induction). Introduce the ratios
\begin{align}
C_n &= \frac{A_n}{A_{n-1}} \label{eq:Cn} \\
D_n &= \frac{B_{n-1}}{B_n} \label{eq:Dn} ,
\end{align}
which obey the recurrence relations (that follow by combining Eqs.~\eqref{eq:Cn} and~\eqref{eq:Dn} with Eqs.~\eqref{eq:An} and~\eqref{eq:Bn})
\begin{align}
C_n &= b_n + \frac{a_n}{C_{n-1}} \\
D_n &= \frac{1}{b_n + a_n D_{n-1}} 
\end{align}
with initial conditions $C_0 = b_0$ and $D_0 = 0$. They are linked to the original $x^{(n)}$ by
\begin{align}
x^{(n)} &= x^{(n-1)} C_n D_n .
\end{align}
$C_n$ and $D_n$ can be computed iteratively starting with the initial value $C_0$ and $D_0$. In each step of the iteration, the new $x^{(n)}$ is computed and the iteration is terminated once the change is below some threshold value, $|x^{(n)} - x^{(n-1)}| = |C_n D_n - 1| < \varepsilon$. In this way, an essentially arbitrary number of $x_m$ (where $m\neq 0,\pm 1$) can be computed. In particular, with $x_{\pm 2}$ known, we can solve the three equations with source terms as discussed in the main text, which gives the components $\delta \mu_{-1}, \delta \mu_0,$ and $\delta \mu_1$. The remaining $\delta \mu_m$ then follow from the simple recursion
\begin{align}
\delta \mu_m &= \begin{cases}
x_m x_{m-1} \ldots x_2 \delta \mu_1 & m\geq 2 \\
x_m x_{m+1} \ldots x_{-2} \delta \mu_{-1} & m\leq -2 .
\end{cases} .
\end{align}

\bibliography{bib_fermiliquid}

\begin{thebibliography}{67}%
\makeatletter
\providecommand \@ifxundefined [1]{%
 \@ifx{#1\undefined}
}%
\providecommand \@ifnum [1]{%
 \ifnum #1\expandafter \@firstoftwo
 \else \expandafter \@secondoftwo
 \fi
}%
\providecommand \@ifx [1]{%
 \ifx #1\expandafter \@firstoftwo
 \else \expandafter \@secondoftwo
 \fi
}%
\providecommand \natexlab [1]{#1}%
\providecommand \enquote  [1]{``#1''}%
\providecommand \bibnamefont  [1]{#1}%
\providecommand \bibfnamefont [1]{#1}%
\providecommand \citenamefont [1]{#1}%
\providecommand \href@noop [0]{\@secondoftwo}%
\providecommand \href [0]{\begingroup \@sanitize@url \@href}%
\providecommand \@href[1]{\@@startlink{#1}\@@href}%
\providecommand \@@href[1]{\endgroup#1\@@endlink}%
\providecommand \@sanitize@url [0]{\catcode `\\12\catcode `\$12\catcode
  `\&12\catcode `\#12\catcode `\^12\catcode `\_12\catcode `\%12\relax}%
\providecommand \@@startlink[1]{}%
\providecommand \@@endlink[0]{}%
\providecommand \url  [0]{\begingroup\@sanitize@url \@url }%
\providecommand \@url [1]{\endgroup\@href {#1}{\urlprefix }}%
\providecommand \urlprefix  [0]{URL }%
\providecommand \Eprint [0]{\href }%
\providecommand \doibase [0]{http://dx.doi.org/}%
\providecommand \selectlanguage [0]{\@gobble}%
\providecommand \bibinfo  [0]{\@secondoftwo}%
\providecommand \bibfield  [0]{\@secondoftwo}%
\providecommand \translation [1]{[#1]}%
\providecommand \BibitemOpen [0]{}%
\providecommand \bibitemStop [0]{}%
\providecommand \bibitemNoStop [0]{.\EOS\space}%
\providecommand \EOS [0]{\spacefactor3000\relax}%
\providecommand \BibitemShut  [1]{\csname bibitem#1\endcsname}%
\let\auto@bib@innerbib\@empty
\bibitem [{\citenamefont {Bandurin}\ \emph {et~al.}(2016)\citenamefont
  {Bandurin}, \citenamefont {Torre}, \citenamefont {Kumar}, \citenamefont
  {Shalom}, \citenamefont {Tomadin}, \citenamefont {Principi}, \citenamefont
  {Auton}, \citenamefont {Khestanova}, \citenamefont {Novoselov}, \citenamefont
  {Grigorieva}, \citenamefont {Ponomarenko}, \citenamefont {Geim},\ and\
  \citenamefont {Polini}}]{bandurin16}%
  \BibitemOpen
  \bibfield  {author} {\bibinfo {author} {\bibfnamefont {D.~A.}\ \bibnamefont
  {Bandurin}}, \bibinfo {author} {\bibfnamefont {I.}~\bibnamefont {Torre}},
  \bibinfo {author} {\bibfnamefont {R.~Krishna}\ \bibnamefont {Kumar}},
  \bibinfo {author} {\bibfnamefont {M.~Ben}\ \bibnamefont {Shalom}}, \bibinfo
  {author} {\bibfnamefont {A.}~\bibnamefont {Tomadin}}, \bibinfo {author}
  {\bibfnamefont {A.}~\bibnamefont {Principi}}, \bibinfo {author}
  {\bibfnamefont {G.~H.}\ \bibnamefont {Auton}}, \bibinfo {author}
  {\bibfnamefont {E.}~\bibnamefont {Khestanova}}, \bibinfo {author}
  {\bibfnamefont {K.~S.}\ \bibnamefont {Novoselov}}, \bibinfo {author}
  {\bibfnamefont {I.~V.}\ \bibnamefont {Grigorieva}}, \bibinfo {author}
  {\bibfnamefont {L.~A.}\ \bibnamefont {Ponomarenko}}, \bibinfo {author}
  {\bibfnamefont {A.~K.}\ \bibnamefont {Geim}}, \ and\ \bibinfo {author}
  {\bibfnamefont {M.}~\bibnamefont {Polini}},\ }\bibfield  {title} {\enquote
  {\bibinfo {title} {{Negative local resistance caused by viscous electron
  backflow in graphene}},}\ }\href {\doibase 10.1126/science.aad0201}
  {\bibfield  {journal} {\bibinfo  {journal} {Science}\ }\textbf {\bibinfo
  {volume} {351}},\ \bibinfo {pages} {1055} (\bibinfo {year}
  {2016})}\BibitemShut {NoStop}%
\bibitem [{\citenamefont {Crossno}\ \emph {et~al.}(2016)\citenamefont
  {Crossno}, \citenamefont {Shi}, \citenamefont {Wang}, \citenamefont {Liu},
  \citenamefont {Harzheim}, \citenamefont {Lucas}, \citenamefont {Sachdev},
  \citenamefont {Kim}, \citenamefont {Taniguchi}, \citenamefont {Watanabe},
  \citenamefont {Ohki},\ and\ \citenamefont {Fong}}]{crossno16}%
  \BibitemOpen
  \bibfield  {author} {\bibinfo {author} {\bibfnamefont {J.}~\bibnamefont
  {Crossno}}, \bibinfo {author} {\bibfnamefont {J.~K.}\ \bibnamefont {Shi}},
  \bibinfo {author} {\bibfnamefont {K.}~\bibnamefont {Wang}}, \bibinfo {author}
  {\bibfnamefont {X.}~\bibnamefont {Liu}}, \bibinfo {author} {\bibfnamefont
  {A.}~\bibnamefont {Harzheim}}, \bibinfo {author} {\bibfnamefont
  {A.}~\bibnamefont {Lucas}}, \bibinfo {author} {\bibfnamefont
  {S.}~\bibnamefont {Sachdev}}, \bibinfo {author} {\bibfnamefont
  {P.}~\bibnamefont {Kim}}, \bibinfo {author} {\bibfnamefont {T.}~\bibnamefont
  {Taniguchi}}, \bibinfo {author} {\bibfnamefont {K.}~\bibnamefont {Watanabe}},
  \bibinfo {author} {\bibfnamefont {T.~A.}\ \bibnamefont {Ohki}}, \ and\
  \bibinfo {author} {\bibfnamefont {K.~C.}\ \bibnamefont {Fong}},\ }\bibfield
  {title} {\enquote {\bibinfo {title} {{Observation of the Dirac fluid and the
  breakdown of the Wiedemann-Franz law in graphene}},}\ }\href {\doibase
  10.1126/science.aad0343} {\bibfield  {journal} {\bibinfo  {journal}
  {Science}\ }\textbf {\bibinfo {volume} {351}},\ \bibinfo {pages} {1058}
  (\bibinfo {year} {2016})}\BibitemShut {NoStop}%
\bibitem [{\citenamefont {Moll}\ \emph {et~al.}(2016)\citenamefont {Moll},
  \citenamefont {Kushwaha}, \citenamefont {Nandi}, \citenamefont {Schmidt},\
  and\ \citenamefont {Mackenzie}}]{moll16}%
  \BibitemOpen
  \bibfield  {author} {\bibinfo {author} {\bibfnamefont {P.~J.~W.}\
  \bibnamefont {Moll}}, \bibinfo {author} {\bibfnamefont {P.}~\bibnamefont
  {Kushwaha}}, \bibinfo {author} {\bibfnamefont {N.}~\bibnamefont {Nandi}},
  \bibinfo {author} {\bibfnamefont {B.}~\bibnamefont {Schmidt}}, \ and\
  \bibinfo {author} {\bibfnamefont {A.~P.}\ \bibnamefont {Mackenzie}},\
  }\bibfield  {title} {\enquote {\bibinfo {title} {{Evidence for hydrodynamic
  electron flow in PdCoO$_2$}},}\ }\href {\doibase 10.1126/science.aac8385}
  {\bibfield  {journal} {\bibinfo  {journal} {Science}\ }\textbf {\bibinfo
  {volume} {351}},\ \bibinfo {pages} {1061} (\bibinfo {year}
  {2016})}\BibitemShut {NoStop}%
\bibitem [{\citenamefont {Nam}\ \emph {et~al.}(2017)\citenamefont {Nam},
  \citenamefont {Ki}, \citenamefont {Soler-Delgado},\ and\ \citenamefont
  {Morpurgo}}]{nam17}%
  \BibitemOpen
  \bibfield  {author} {\bibinfo {author} {\bibfnamefont {Y.}~\bibnamefont
  {Nam}}, \bibinfo {author} {\bibfnamefont {D.-K.}\ \bibnamefont {Ki}},
  \bibinfo {author} {\bibfnamefont {D.}~\bibnamefont {Soler-Delgado}}, \ and\
  \bibinfo {author} {\bibfnamefont {A.~F.}\ \bibnamefont {Morpurgo}},\
  }\bibfield  {title} {\enquote {\bibinfo {title} {{Electron--hole collision
  limited transport in charge-neutral bilayer graphene}},}\ }\href {\doibase
  10.1038/nphys4218} {\bibfield  {journal} {\bibinfo  {journal} {Nature
  Physics}\ }\textbf {\bibinfo {volume} {13}},\ \bibinfo {pages} {1207}
  (\bibinfo {year} {2017})}\BibitemShut {NoStop}%
\bibitem [{\citenamefont {Krishna~Kumar}\ \emph {et~al.}(2017)\citenamefont
  {Krishna~Kumar}, \citenamefont {Bandurin}, \citenamefont {Pellegrino},
  \citenamefont {Cao}, \citenamefont {Principi}, \citenamefont {Guo},
  \citenamefont {Auton}, \citenamefont {Ben~Shalom}, \citenamefont
  {Ponomarenko}, \citenamefont {Falkovich}, \citenamefont {Watanabe},
  \citenamefont {Taniguchi}, \citenamefont {Grigorieva}, \citenamefont
  {Levitov}, \citenamefont {Polini},\ and\ \citenamefont
  {Geim}}]{krishnakumar17}%
  \BibitemOpen
  \bibfield  {author} {\bibinfo {author} {\bibfnamefont {R.}~\bibnamefont
  {Krishna~Kumar}}, \bibinfo {author} {\bibfnamefont {D.~A.}\ \bibnamefont
  {Bandurin}}, \bibinfo {author} {\bibfnamefont {F.~M.~D.}\ \bibnamefont
  {Pellegrino}}, \bibinfo {author} {\bibfnamefont {Y.}~\bibnamefont {Cao}},
  \bibinfo {author} {\bibfnamefont {A.}~\bibnamefont {Principi}}, \bibinfo
  {author} {\bibfnamefont {H.}~\bibnamefont {Guo}}, \bibinfo {author}
  {\bibfnamefont {G.~H.}\ \bibnamefont {Auton}}, \bibinfo {author}
  {\bibfnamefont {M.}~\bibnamefont {Ben~Shalom}}, \bibinfo {author}
  {\bibfnamefont {L.~A.}\ \bibnamefont {Ponomarenko}}, \bibinfo {author}
  {\bibfnamefont {G.}~\bibnamefont {Falkovich}}, \bibinfo {author}
  {\bibfnamefont {K.}~\bibnamefont {Watanabe}}, \bibinfo {author}
  {\bibfnamefont {T.}~\bibnamefont {Taniguchi}}, \bibinfo {author}
  {\bibfnamefont {I.~V.}\ \bibnamefont {Grigorieva}}, \bibinfo {author}
  {\bibfnamefont {L.~S.}\ \bibnamefont {Levitov}}, \bibinfo {author}
  {\bibfnamefont {M.}~\bibnamefont {Polini}}, \ and\ \bibinfo {author}
  {\bibfnamefont {A.~K.}\ \bibnamefont {Geim}},\ }\bibfield  {title} {\enquote
  {\bibinfo {title} {{Superballistic flow of viscous electron fluid through
  graphene constrictions}},}\ }\href {\doibase 10.1038/nphys4240} {\bibfield
  {journal} {\bibinfo  {journal} {Nature Physics}\ }\textbf {\bibinfo {volume}
  {13}},\ \bibinfo {pages} {1182} (\bibinfo {year} {2017})}\BibitemShut
  {NoStop}%
\bibitem [{\citenamefont {Gooth}\ \emph {et~al.}(2018)\citenamefont {Gooth},
  \citenamefont {Menges}, \citenamefont {Kumar}, \citenamefont {S\"{u}\ss},
  \citenamefont {Shekhar}, \citenamefont {Sun}, \citenamefont {Drechsler},
  \citenamefont {Zierold}, \citenamefont {Felser},\ and\ \citenamefont
  {Gotsmann}}]{gooth18}%
  \BibitemOpen
  \bibfield  {author} {\bibinfo {author} {\bibfnamefont {J.}~\bibnamefont
  {Gooth}}, \bibinfo {author} {\bibfnamefont {F.}~\bibnamefont {Menges}},
  \bibinfo {author} {\bibfnamefont {N.}~\bibnamefont {Kumar}}, \bibinfo
  {author} {\bibfnamefont {V.}~\bibnamefont {S\"{u}\ss}}, \bibinfo {author}
  {\bibfnamefont {C.}~\bibnamefont {Shekhar}}, \bibinfo {author} {\bibfnamefont
  {Y.}~\bibnamefont {Sun}}, \bibinfo {author} {\bibfnamefont {U.}~\bibnamefont
  {Drechsler}}, \bibinfo {author} {\bibfnamefont {R.}~\bibnamefont {Zierold}},
  \bibinfo {author} {\bibfnamefont {C.}~\bibnamefont {Felser}}, \ and\ \bibinfo
  {author} {\bibfnamefont {B.}~\bibnamefont {Gotsmann}},\ }\bibfield  {title}
  {\enquote {\bibinfo {title} {{Thermal and electrical signatures of a
  hydrodynamic electron fluid in tungsten diphosphide}},}\ }\href {\doibase
  10.1038/s41467-018-06688-y} {\bibfield  {journal} {\bibinfo  {journal}
  {Nature Communications}\ }\textbf {\bibinfo {volume} {9}},\ \bibinfo {pages}
  {4093} (\bibinfo {year} {2018})}\BibitemShut {NoStop}%
\bibitem [{\citenamefont {Gusev}\ \emph {et~al.}(2018)\citenamefont {Gusev},
  \citenamefont {Levin}, \citenamefont {Levinson},\ and\ \citenamefont
  {Bakarov}}]{gusev18}%
  \BibitemOpen
  \bibfield  {author} {\bibinfo {author} {\bibfnamefont {G.~M.}\ \bibnamefont
  {Gusev}}, \bibinfo {author} {\bibfnamefont {A.~D.}\ \bibnamefont {Levin}},
  \bibinfo {author} {\bibfnamefont {E.~V.}\ \bibnamefont {Levinson}}, \ and\
  \bibinfo {author} {\bibfnamefont {A.~K.}\ \bibnamefont {Bakarov}},\
  }\bibfield  {title} {\enquote {\bibinfo {title} {{Viscous transport and Hall
  viscosity in a two-dimensional electron system}},}\ }\href {\doibase
  10.1103/PhysRevB.98.161303} {\bibfield  {journal} {\bibinfo  {journal} {Phys.
  Rev. B}\ }\textbf {\bibinfo {volume} {98}},\ \bibinfo {pages} {161303}
  (\bibinfo {year} {2018})}\BibitemShut {NoStop}%
\bibitem [{\citenamefont {Bandurin}\ \emph {et~al.}(2018)\citenamefont
  {Bandurin}, \citenamefont {Shytov}, \citenamefont {Levitov}, \citenamefont
  {Kumar}, \citenamefont {Berdyugin}, \citenamefont {Ben~Shalom}, \citenamefont
  {Grigorieva}, \citenamefont {Geim},\ and\ \citenamefont
  {Falkovich}}]{bandurin18}%
  \BibitemOpen
  \bibfield  {author} {\bibinfo {author} {\bibfnamefont {D.~A.}\ \bibnamefont
  {Bandurin}}, \bibinfo {author} {\bibfnamefont {A.~V.}\ \bibnamefont
  {Shytov}}, \bibinfo {author} {\bibfnamefont {L.~S.}\ \bibnamefont {Levitov}},
  \bibinfo {author} {\bibfnamefont {R.~K.}\ \bibnamefont {Kumar}}, \bibinfo
  {author} {\bibfnamefont {A.~I.}\ \bibnamefont {Berdyugin}}, \bibinfo {author}
  {\bibfnamefont {M.}~\bibnamefont {Ben~Shalom}}, \bibinfo {author}
  {\bibfnamefont {I.~V.}\ \bibnamefont {Grigorieva}}, \bibinfo {author}
  {\bibfnamefont {A.~K.}\ \bibnamefont {Geim}}, \ and\ \bibinfo {author}
  {\bibfnamefont {G.}~\bibnamefont {Falkovich}},\ }\bibfield  {title} {\enquote
  {\bibinfo {title} {{Fluidity onset in graphene}},}\ }\href {\doibase
  10.1038/s41467-018-07004-4} {\bibfield  {journal} {\bibinfo  {journal}
  {Nature Communications}\ }\textbf {\bibinfo {volume} {9}},\ \bibinfo {pages}
  {4533} (\bibinfo {year} {2018})}\BibitemShut {NoStop}%
\bibitem [{\citenamefont {Braem}\ \emph {et~al.}(2018)\citenamefont {Braem},
  \citenamefont {Pellegrino}, \citenamefont {Principi}, \citenamefont
  {R\"o\"osli}, \citenamefont {Gold}, \citenamefont {Hennel}, \citenamefont
  {Koski}, \citenamefont {Berl}, \citenamefont {Dietsche}, \citenamefont
  {Wegscheider}, \citenamefont {Polini}, \citenamefont {Ihn},\ and\
  \citenamefont {Ensslin}}]{braem18}%
  \BibitemOpen
  \bibfield  {author} {\bibinfo {author} {\bibfnamefont {B.~A.}\ \bibnamefont
  {Braem}}, \bibinfo {author} {\bibfnamefont {F.~M.~D.}\ \bibnamefont
  {Pellegrino}}, \bibinfo {author} {\bibfnamefont {A.}~\bibnamefont
  {Principi}}, \bibinfo {author} {\bibfnamefont {M.}~\bibnamefont
  {R\"o\"osli}}, \bibinfo {author} {\bibfnamefont {C.}~\bibnamefont {Gold}},
  \bibinfo {author} {\bibfnamefont {S.}~\bibnamefont {Hennel}}, \bibinfo
  {author} {\bibfnamefont {J.~V.}\ \bibnamefont {Koski}}, \bibinfo {author}
  {\bibfnamefont {M.}~\bibnamefont {Berl}}, \bibinfo {author} {\bibfnamefont
  {W.}~\bibnamefont {Dietsche}}, \bibinfo {author} {\bibfnamefont
  {W.}~\bibnamefont {Wegscheider}}, \bibinfo {author} {\bibfnamefont
  {M.}~\bibnamefont {Polini}}, \bibinfo {author} {\bibfnamefont
  {T.}~\bibnamefont {Ihn}}, \ and\ \bibinfo {author} {\bibfnamefont
  {K.}~\bibnamefont {Ensslin}},\ }\bibfield  {title} {\enquote {\bibinfo
  {title} {{Scanning gate microscopy in a viscous electron fluid}},}\ }\href
  {\doibase 10.1103/PhysRevB.98.241304} {\bibfield  {journal} {\bibinfo
  {journal} {Phys. Rev. B}\ }\textbf {\bibinfo {volume} {98}},\ \bibinfo
  {pages} {241304} (\bibinfo {year} {2018})}\BibitemShut {NoStop}%
\bibitem [{\citenamefont {Berdyugin}\ \emph {et~al.}(2019)\citenamefont
  {Berdyugin}, \citenamefont {Xu}, \citenamefont {Pellegrino}, \citenamefont
  {Kumar}, \citenamefont {Principi}, \citenamefont {Torre}, \citenamefont
  {Shalom}, \citenamefont {Taniguchi}, \citenamefont {Watanabe}, \citenamefont
  {Grigorieva}, \citenamefont {Polini}, \citenamefont {Geim},\ and\
  \citenamefont {Bandurin}}]{berdyugin19}%
  \BibitemOpen
  \bibfield  {author} {\bibinfo {author} {\bibfnamefont {A.~I.}\ \bibnamefont
  {Berdyugin}}, \bibinfo {author} {\bibfnamefont {S.~G.}\ \bibnamefont {Xu}},
  \bibinfo {author} {\bibfnamefont {F.~M.~D.}\ \bibnamefont {Pellegrino}},
  \bibinfo {author} {\bibfnamefont {R.~Krishna}\ \bibnamefont {Kumar}},
  \bibinfo {author} {\bibfnamefont {A.}~\bibnamefont {Principi}}, \bibinfo
  {author} {\bibfnamefont {I.}~\bibnamefont {Torre}}, \bibinfo {author}
  {\bibfnamefont {M.~Ben}\ \bibnamefont {Shalom}}, \bibinfo {author}
  {\bibfnamefont {T.}~\bibnamefont {Taniguchi}}, \bibinfo {author}
  {\bibfnamefont {K.}~\bibnamefont {Watanabe}}, \bibinfo {author}
  {\bibfnamefont {I.~V.}\ \bibnamefont {Grigorieva}}, \bibinfo {author}
  {\bibfnamefont {M.}~\bibnamefont {Polini}}, \bibinfo {author} {\bibfnamefont
  {A.~K.}\ \bibnamefont {Geim}}, \ and\ \bibinfo {author} {\bibfnamefont
  {D.~A.}\ \bibnamefont {Bandurin}},\ }\bibfield  {title} {\enquote {\bibinfo
  {title} {{Measuring Hall viscosity of graphene's electron fluid}},}\ }\href
  {\doibase 10.1126/science.aau0685} {\bibfield  {journal} {\bibinfo  {journal}
  {Science}\ }\textbf {\bibinfo {volume} {364}},\ \bibinfo {pages} {162}
  (\bibinfo {year} {2019})}\BibitemShut {NoStop}%
\bibitem [{\citenamefont {Gallagher}\ \emph {et~al.}(2019)\citenamefont
  {Gallagher}, \citenamefont {Yang}, \citenamefont {Lyu}, \citenamefont {Tian},
  \citenamefont {Kou}, \citenamefont {Zhang}, \citenamefont {Watanabe},
  \citenamefont {Taniguchi},\ and\ \citenamefont {Wang}}]{gallagher19}%
  \BibitemOpen
  \bibfield  {author} {\bibinfo {author} {\bibfnamefont {P.}~\bibnamefont
  {Gallagher}}, \bibinfo {author} {\bibfnamefont {C.-S.}\ \bibnamefont {Yang}},
  \bibinfo {author} {\bibfnamefont {T.}~\bibnamefont {Lyu}}, \bibinfo {author}
  {\bibfnamefont {F.}~\bibnamefont {Tian}}, \bibinfo {author} {\bibfnamefont
  {R.}~\bibnamefont {Kou}}, \bibinfo {author} {\bibfnamefont {H.}~\bibnamefont
  {Zhang}}, \bibinfo {author} {\bibfnamefont {K.}~\bibnamefont {Watanabe}},
  \bibinfo {author} {\bibfnamefont {T.}~\bibnamefont {Taniguchi}}, \ and\
  \bibinfo {author} {\bibfnamefont {F.}~\bibnamefont {Wang}},\ }\bibfield
  {title} {\enquote {\bibinfo {title} {{Quantum-critical conductivity of the
  Dirac fluid in graphene}},}\ }\href {\doibase 10.1126/science.aat8687}
  {\bibfield  {journal} {\bibinfo  {journal} {Science}\ }\textbf {\bibinfo
  {volume} {364}},\ \bibinfo {pages} {158} (\bibinfo {year}
  {2019})}\BibitemShut {NoStop}%
\bibitem [{\citenamefont {Sulpizio}\ \emph {et~al.}(2019)\citenamefont
  {Sulpizio}, \citenamefont {Ella}, \citenamefont {Rozen}, \citenamefont
  {Birkbeck}, \citenamefont {Perello}, \citenamefont {Dutta}, \citenamefont
  {Ben-Shalom}, \citenamefont {Taniguchi}, \citenamefont {Watanabe},
  \citenamefont {Holder}, \citenamefont {Queiroz}, \citenamefont {Principi},
  \citenamefont {Stern}, \citenamefont {Scaffidi}, \citenamefont {Geim},\ and\
  \citenamefont {Ilani}}]{sulpizio19}%
  \BibitemOpen
  \bibfield  {author} {\bibinfo {author} {\bibfnamefont {J.~A.}\ \bibnamefont
  {Sulpizio}}, \bibinfo {author} {\bibfnamefont {L.}~\bibnamefont {Ella}},
  \bibinfo {author} {\bibfnamefont {A.}~\bibnamefont {Rozen}}, \bibinfo
  {author} {\bibfnamefont {J.}~\bibnamefont {Birkbeck}}, \bibinfo {author}
  {\bibfnamefont {D.~J.}\ \bibnamefont {Perello}}, \bibinfo {author}
  {\bibfnamefont {D.}~\bibnamefont {Dutta}}, \bibinfo {author} {\bibfnamefont
  {M.}~\bibnamefont {Ben-Shalom}}, \bibinfo {author} {\bibfnamefont
  {T.}~\bibnamefont {Taniguchi}}, \bibinfo {author} {\bibfnamefont
  {K.}~\bibnamefont {Watanabe}}, \bibinfo {author} {\bibfnamefont
  {T.}~\bibnamefont {Holder}}, \bibinfo {author} {\bibfnamefont
  {R.}~\bibnamefont {Queiroz}}, \bibinfo {author} {\bibfnamefont
  {A.}~\bibnamefont {Principi}}, \bibinfo {author} {\bibfnamefont
  {A.}~\bibnamefont {Stern}}, \bibinfo {author} {\bibfnamefont
  {T.}~\bibnamefont {Scaffidi}}, \bibinfo {author} {\bibfnamefont {A.~K.}\
  \bibnamefont {Geim}}, \ and\ \bibinfo {author} {\bibfnamefont
  {S.}~\bibnamefont {Ilani}},\ }\bibfield  {title} {\enquote {\bibinfo {title}
  {{Visualizing Poiseuille flow of hydrodynamic electrons}},}\ }\href {\doibase
  10.1038/s41586-019-1788-9} {\bibfield  {journal} {\bibinfo  {journal}
  {Nature}\ }\textbf {\bibinfo {volume} {576}},\ \bibinfo {pages} {75}
  (\bibinfo {year} {2019})}\BibitemShut {NoStop}%
\bibitem [{\citenamefont {Jenkins}\ \emph {et~al.}(2022)\citenamefont
  {Jenkins}, \citenamefont {Baumann}, \citenamefont {Zhou}, \citenamefont
  {Meynell}, \citenamefont {Daipeng}, \citenamefont {Watanabe}, \citenamefont
  {Taniguchi}, \citenamefont {Lucas}, \citenamefont {Young},\ and\
  \citenamefont {Bleszynski~Jayich}}]{jenkins22}%
  \BibitemOpen
  \bibfield  {author} {\bibinfo {author} {\bibfnamefont {A.}~\bibnamefont
  {Jenkins}}, \bibinfo {author} {\bibfnamefont {S.}~\bibnamefont {Baumann}},
  \bibinfo {author} {\bibfnamefont {H.}~\bibnamefont {Zhou}}, \bibinfo {author}
  {\bibfnamefont {S.~A.}\ \bibnamefont {Meynell}}, \bibinfo {author}
  {\bibfnamefont {Y.}~\bibnamefont {Daipeng}}, \bibinfo {author} {\bibfnamefont
  {K.}~\bibnamefont {Watanabe}}, \bibinfo {author} {\bibfnamefont
  {T.}~\bibnamefont {Taniguchi}}, \bibinfo {author} {\bibfnamefont
  {A.}~\bibnamefont {Lucas}}, \bibinfo {author} {\bibfnamefont {A.~F.}\
  \bibnamefont {Young}}, \ and\ \bibinfo {author} {\bibfnamefont {A.~C.}\
  \bibnamefont {Bleszynski~Jayich}},\ }\bibfield  {title} {\enquote {\bibinfo
  {title} {{Imaging the Breakdown of Ohmic Transport in Graphene}},}\ }\href
  {\doibase 10.1103/PhysRevLett.129.087701} {\bibfield  {journal} {\bibinfo
  {journal} {Phys. Rev. Lett.}\ }\textbf {\bibinfo {volume} {129}},\ \bibinfo
  {pages} {087701} (\bibinfo {year} {2022})}\BibitemShut {NoStop}%
\bibitem [{\citenamefont {de~Jong}\ and\ \citenamefont
  {Molenkamp}(1995)}]{dejong95}%
  \BibitemOpen
  \bibfield  {author} {\bibinfo {author} {\bibfnamefont {M.~J.~M.}\
  \bibnamefont {de~Jong}}\ and\ \bibinfo {author} {\bibfnamefont {L.~W.}\
  \bibnamefont {Molenkamp}},\ }\bibfield  {title} {\enquote {\bibinfo {title}
  {{Hydrodynamic electron flow in high-mobility wires}},}\ }\href {\doibase
  10.1103/PhysRevB.51.13389} {\bibfield  {journal} {\bibinfo  {journal} {Phys.
  Rev. B}\ }\textbf {\bibinfo {volume} {51}},\ \bibinfo {pages} {13389}
  (\bibinfo {year} {1995})}\BibitemShut {NoStop}%
\bibitem [{\citenamefont {Buhmann}\ and\ \citenamefont
  {Molenkamp}(2002)}]{buhmann02}%
  \BibitemOpen
  \bibfield  {author} {\bibinfo {author} {\bibfnamefont {H.}~\bibnamefont
  {Buhmann}}\ and\ \bibinfo {author} {\bibfnamefont {L.W.}\ \bibnamefont
  {Molenkamp}},\ }\bibfield  {title} {\enquote {\bibinfo {title} {{1D
  diffusion: a novel transport regime in narrow 2DEG channels}},}\ }\href
  {\doibase https://doi.org/10.1016/S1386-9477(01)00387-3} {\bibfield
  {journal} {\bibinfo  {journal} {Physica E}\ }\textbf {\bibinfo {volume}
  {12}},\ \bibinfo {pages} {715} (\bibinfo {year} {2002})}\BibitemShut
  {NoStop}%
\bibitem [{\citenamefont {Baym}\ and\ \citenamefont {Pethick}(2004)}]{baym04}%
  \BibitemOpen
  \bibfield  {author} {\bibinfo {author} {\bibfnamefont {G.}~\bibnamefont
  {Baym}}\ and\ \bibinfo {author} {\bibfnamefont {C.}~\bibnamefont {Pethick}},\
  }\href@noop {} {\emph {\bibinfo {title} {{Landau Fermi-Liquid Theory}}}}\
  (\bibinfo  {publisher} {Wiley-VCH Verlag (Weinheim)},\ \bibinfo {year}
  {2004})\BibitemShut {NoStop}%
\bibitem [{\citenamefont {Giuliani}\ and\ \citenamefont
  {Vignale}(2005)}]{giuliani05}%
  \BibitemOpen
  \bibfield  {author} {\bibinfo {author} {\bibfnamefont {G.~F.}\ \bibnamefont
  {Giuliani}}\ and\ \bibinfo {author} {\bibfnamefont {G.}~\bibnamefont
  {Vignale}},\ }\href@noop {} {\emph {\bibinfo {title} {{Quantum Theory of the
  Electron Liquid}}}}\ (\bibinfo  {publisher} {Cambridge University Press
  (Cambridge)},\ \bibinfo {year} {2005})\BibitemShut {NoStop}%
\bibitem [{\citenamefont {Pines}\ and\ \citenamefont
  {Nozi\`{e}res}(2018)}]{pines18}%
  \BibitemOpen
  \bibfield  {author} {\bibinfo {author} {\bibfnamefont {D.}~\bibnamefont
  {Pines}}\ and\ \bibinfo {author} {\bibfnamefont {P.}~\bibnamefont
  {Nozi\`{e}res}},\ }\href@noop {} {\emph {\bibinfo {title} {{The Theory of
  Quantum Liquids}}}}\ (\bibinfo  {publisher} {CRC Press (Boca Raton)},\
  \bibinfo {year} {2018})\BibitemShut {NoStop}%
\bibitem [{\citenamefont {Giuliani}\ and\ \citenamefont
  {Quinn}(1982)}]{giuliani82}%
  \BibitemOpen
  \bibfield  {author} {\bibinfo {author} {\bibfnamefont {G.~F.}\ \bibnamefont
  {Giuliani}}\ and\ \bibinfo {author} {\bibfnamefont {J.~J.}\ \bibnamefont
  {Quinn}},\ }\bibfield  {title} {\enquote {\bibinfo {title} {{Lifetime of a
  quasiparticle in a two-dimensional electron gas}},}\ }\href {\doibase
  10.1103/PhysRevB.26.4421} {\bibfield  {journal} {\bibinfo  {journal} {Phys.
  Rev. B}\ }\textbf {\bibinfo {volume} {26}},\ \bibinfo {pages} {4421}
  (\bibinfo {year} {1982})}\BibitemShut {NoStop}%
\bibitem [{\citenamefont {Zheng}\ and\ \citenamefont
  {Das~Sarma}(1996)}]{zheng96}%
  \BibitemOpen
  \bibfield  {author} {\bibinfo {author} {\bibfnamefont {L.}~\bibnamefont
  {Zheng}}\ and\ \bibinfo {author} {\bibfnamefont {S.}~\bibnamefont
  {Das~Sarma}},\ }\bibfield  {title} {\enquote {\bibinfo {title} {{Coulomb
  scattering lifetime of a two-dimensional electron gas}},}\ }\href {\doibase
  10.1103/PhysRevB.53.9964} {\bibfield  {journal} {\bibinfo  {journal} {Phys.
  Rev. B}\ }\textbf {\bibinfo {volume} {53}},\ \bibinfo {pages} {9964}
  (\bibinfo {year} {1996})}\BibitemShut {NoStop}%
\bibitem [{\citenamefont {Li}\ and\ \citenamefont {Das~Sarma}(2013)}]{li13}%
  \BibitemOpen
  \bibfield  {author} {\bibinfo {author} {\bibfnamefont {Q.}~\bibnamefont
  {Li}}\ and\ \bibinfo {author} {\bibfnamefont {S.}~\bibnamefont {Das~Sarma}},\
  }\bibfield  {title} {\enquote {\bibinfo {title} {{Finite temperature
  inelastic mean free path and quasiparticle lifetime in graphene}},}\ }\href
  {\doibase 10.1103/PhysRevB.87.085406} {\bibfield  {journal} {\bibinfo
  {journal} {Phys. Rev. B}\ }\textbf {\bibinfo {volume} {87}},\ \bibinfo
  {pages} {085406} (\bibinfo {year} {2013})}\BibitemShut {NoStop}%
\bibitem [{\citenamefont {{Das Sarma}}\ and\ \citenamefont
  {Liao}(2021)}]{dassarma21}%
  \BibitemOpen
  \bibfield  {author} {\bibinfo {author} {\bibfnamefont {S.}~\bibnamefont {{Das
  Sarma}}}\ and\ \bibinfo {author} {\bibfnamefont {Y.}~\bibnamefont {Liao}},\
  }\bibfield  {title} {\enquote {\bibinfo {title} {{Know the enemy: 2D Fermi
  liquids}},}\ }\href {\doibase https://doi.org/10.1016/j.aop.2021.168495}
  {\bibfield  {journal} {\bibinfo  {journal} {Annals of Physics}\ }\textbf
  {\bibinfo {volume} {435}},\ \bibinfo {pages} {168495} (\bibinfo {year}
  {2021})}\BibitemShut {NoStop}%
\bibitem [{\citenamefont {Ahn}\ and\ \citenamefont {Das~Sarma}(2022)}]{ahn22}%
  \BibitemOpen
  \bibfield  {author} {\bibinfo {author} {\bibfnamefont {S.}~\bibnamefont
  {Ahn}}\ and\ \bibinfo {author} {\bibfnamefont {S.}~\bibnamefont
  {Das~Sarma}},\ }\bibfield  {title} {\enquote {\bibinfo {title}
  {{Hydrodynamics, viscous electron fluid, and Wiedeman-Franz law in
  two-dimensional semiconductors}},}\ }\href {\doibase
  10.1103/PhysRevB.106.L081303} {\bibfield  {journal} {\bibinfo  {journal}
  {Phys. Rev. B}\ }\textbf {\bibinfo {volume} {106}},\ \bibinfo {pages}
  {L081303} (\bibinfo {year} {2022})}\BibitemShut {NoStop}%
\bibitem [{\citenamefont {Laikhtman}(1992)}]{laikhtman92}%
  \BibitemOpen
  \bibfield  {author} {\bibinfo {author} {\bibfnamefont {B.}~\bibnamefont
  {Laikhtman}},\ }\bibfield  {title} {\enquote {\bibinfo {title}
  {{Electron-electron angular relaxation in a two-dimensional electron gas}},}\
  }\href {\doibase 10.1103/PhysRevB.45.1259} {\bibfield  {journal} {\bibinfo
  {journal} {Phys. Rev. B}\ }\textbf {\bibinfo {volume} {45}},\ \bibinfo
  {pages} {1259} (\bibinfo {year} {1992})}\BibitemShut {NoStop}%
\bibitem [{\citenamefont {Gurzhi}\ \emph {et~al.}(1995)\citenamefont {Gurzhi},
  \citenamefont {Kalinenko},\ and\ \citenamefont {Kopeliovich}}]{gurzhi95}%
  \BibitemOpen
  \bibfield  {author} {\bibinfo {author} {\bibfnamefont {R.~N.}\ \bibnamefont
  {Gurzhi}}, \bibinfo {author} {\bibfnamefont {A.~N.}\ \bibnamefont
  {Kalinenko}}, \ and\ \bibinfo {author} {\bibfnamefont {A.~I.}\ \bibnamefont
  {Kopeliovich}},\ }\bibfield  {title} {\enquote {\bibinfo {title}
  {{Electron-electron momentum relaxation in a two-dimensional electron
  gas}},}\ }\href {\doibase 10.1103/PhysRevB.52.4744} {\bibfield  {journal}
  {\bibinfo  {journal} {Phys. Rev. B}\ }\textbf {\bibinfo {volume} {52}},\
  \bibinfo {pages} {4744} (\bibinfo {year} {1995})}\BibitemShut {NoStop}%
\bibitem [{\citenamefont {Ledwith}\ \emph {et~al.}(2017)\citenamefont
  {Ledwith}, \citenamefont {Guo},\ and\ \citenamefont {Levitov}}]{ledwith17}%
  \BibitemOpen
  \bibfield  {author} {\bibinfo {author} {\bibfnamefont {P.}~\bibnamefont
  {Ledwith}}, \bibinfo {author} {\bibfnamefont {H.}~\bibnamefont {Guo}}, \ and\
  \bibinfo {author} {\bibfnamefont {L.}~\bibnamefont {Levitov}},\ }\bibfield
  {title} {\enquote {\bibinfo {title} {{Angular Superdiffusion and Directional
  Memory in Two-Dimensional Electron Fluids}},}\ }\href@noop {} {\bibfield
  {journal} {\bibinfo  {journal} {arXiv:1708.01915}\ } (\bibinfo {year}
  {2017})}\BibitemShut {NoStop}%
\bibitem [{\citenamefont {Ledwith}\ \emph
  {et~al.}(2019{\natexlab{a}})\citenamefont {Ledwith}, \citenamefont {Guo},\
  and\ \citenamefont {Levitov}}]{ledwith19}%
  \BibitemOpen
  \bibfield  {author} {\bibinfo {author} {\bibfnamefont {P.~J.}\ \bibnamefont
  {Ledwith}}, \bibinfo {author} {\bibfnamefont {H.}~\bibnamefont {Guo}}, \ and\
  \bibinfo {author} {\bibfnamefont {L.}~\bibnamefont {Levitov}},\ }\bibfield
  {title} {\enquote {\bibinfo {title} {{The hierarchy of excitation lifetimes
  in two-dimensional Fermi gases}},}\ }\href {\doibase
  https://doi.org/10.1016/j.aop.2019.167913} {\bibfield  {journal} {\bibinfo
  {journal} {Annals of Physics}\ }\textbf {\bibinfo {volume} {411}},\ \bibinfo
  {pages} {167913} (\bibinfo {year} {2019}{\natexlab{a}})}\BibitemShut
  {NoStop}%
\bibitem [{\citenamefont {Ledwith}\ \emph
  {et~al.}(2019{\natexlab{b}})\citenamefont {Ledwith}, \citenamefont {Guo},
  \citenamefont {Shytov},\ and\ \citenamefont {Levitov}}]{ledwith19b}%
  \BibitemOpen
  \bibfield  {author} {\bibinfo {author} {\bibfnamefont {P.}~\bibnamefont
  {Ledwith}}, \bibinfo {author} {\bibfnamefont {H.}~\bibnamefont {Guo}},
  \bibinfo {author} {\bibfnamefont {A.}~\bibnamefont {Shytov}}, \ and\ \bibinfo
  {author} {\bibfnamefont {L.}~\bibnamefont {Levitov}},\ }\bibfield  {title}
  {\enquote {\bibinfo {title} {{Tomographic Dynamics and Scale-Dependent
  Viscosity in 2D Electron Systems}},}\ }\href {\doibase
  10.1103/PhysRevLett.123.116601} {\bibfield  {journal} {\bibinfo  {journal}
  {Phys. Rev. Lett.}\ }\textbf {\bibinfo {volume} {123}},\ \bibinfo {pages}
  {116601} (\bibinfo {year} {2019}{\natexlab{b}})}\BibitemShut {NoStop}%
\bibitem [{\citenamefont {Hofmann}\ and\ \citenamefont
  {Gran}(2022)}]{hofmann22}%
  \BibitemOpen
  \bibfield  {author} {\bibinfo {author} {\bibfnamefont {J.}~\bibnamefont
  {Hofmann}}\ and\ \bibinfo {author} {\bibfnamefont {U.}~\bibnamefont {Gran}},\
  }\bibfield  {title} {\enquote {\bibinfo {title} {{Anomalously long lifetimes
  in two-dimensional Fermi liquids}},}\ }\href
  {https://arxiv.org/abs/2210.16300} {\bibfield  {journal} {\bibinfo  {journal}
  {arXiv:2210.16300}\ } (\bibinfo {year} {2022})}\BibitemShut {NoStop}%
\bibitem [{\citenamefont {Nilsson}\ and\ \citenamefont
  {Castro~Neto}(2005)}]{nilsson05}%
  \BibitemOpen
  \bibfield  {author} {\bibinfo {author} {\bibfnamefont {J.}~\bibnamefont
  {Nilsson}}\ and\ \bibinfo {author} {\bibfnamefont {A.~H.}\ \bibnamefont
  {Castro~Neto}},\ }\bibfield  {title} {\enquote {\bibinfo {title} {{Heat bath
  approach to Landau damping and Pomeranchuk quantum critical points}},}\
  }\href {\doibase 10.1103/PhysRevB.72.195104} {\bibfield  {journal} {\bibinfo
  {journal} {Phys. Rev. B}\ }\textbf {\bibinfo {volume} {72}},\ \bibinfo
  {pages} {195104} (\bibinfo {year} {2005})}\BibitemShut {NoStop}%
\bibitem [{\citenamefont {Hong}\ \emph {et~al.}(2020)\citenamefont {Hong},
  \citenamefont {Davydova}, \citenamefont {Ledwith},\ and\ \citenamefont
  {Levitov}}]{hong20}%
  \BibitemOpen
  \bibfield  {author} {\bibinfo {author} {\bibfnamefont {Q.}~\bibnamefont
  {Hong}}, \bibinfo {author} {\bibfnamefont {N.}~\bibnamefont {Davydova}},
  \bibinfo {author} {\bibfnamefont {P.~J}\ \bibnamefont {Ledwith}}, \ and\
  \bibinfo {author} {\bibfnamefont {L.}~\bibnamefont {Levitov}},\ }\bibfield
  {title} {\enquote {\bibinfo {title} {{Superscreening by a Retroreflected Hole
  Backflow in Tomographic Electron Fluids}},}\ }\href@noop {} {\bibfield
  {journal} {\bibinfo  {journal} {arxiv:2012.03840}\ } (\bibinfo {year}
  {2020})}\BibitemShut {NoStop}%
\bibitem [{\citenamefont {Kiselev}\ and\ \citenamefont
  {Schmalian}(2019)}]{kiselev19}%
  \BibitemOpen
  \bibfield  {author} {\bibinfo {author} {\bibfnamefont {E.~I.}\ \bibnamefont
  {Kiselev}}\ and\ \bibinfo {author} {\bibfnamefont {J.}~\bibnamefont
  {Schmalian}},\ }\bibfield  {title} {\enquote {\bibinfo {title} {{L\'evy
  Flights and Hydrodynamic Superdiffusion on the Dirac Cone of Graphene}},}\
  }\href {\doibase 10.1103/PhysRevLett.123.195302} {\bibfield  {journal}
  {\bibinfo  {journal} {Phys. Rev. Lett.}\ }\textbf {\bibinfo {volume} {123}},\
  \bibinfo {pages} {195302} (\bibinfo {year} {2019})}\BibitemShut {NoStop}%
\bibitem [{\citenamefont {Kiselev}\ and\ \citenamefont
  {Schmalian}(2020)}]{kiselev20}%
  \BibitemOpen
  \bibfield  {author} {\bibinfo {author} {\bibfnamefont {E.~I.}\ \bibnamefont
  {Kiselev}}\ and\ \bibinfo {author} {\bibfnamefont {J.}~\bibnamefont
  {Schmalian}},\ }\bibfield  {title} {\enquote {\bibinfo {title} {{Nonlocal
  hydrodynamic transport and collective excitations in Dirac fluids}},}\ }\href
  {\doibase 10.1103/PhysRevB.102.245434} {\bibfield  {journal} {\bibinfo
  {journal} {Phys. Rev. B}\ }\textbf {\bibinfo {volume} {102}},\ \bibinfo
  {pages} {245434} (\bibinfo {year} {2020})}\BibitemShut {NoStop}%
\bibitem [{\citenamefont {Landau}(1957{\natexlab{a}})}]{landau57a}%
  \BibitemOpen
  \bibfield  {author} {\bibinfo {author} {\bibfnamefont {L.~D.}\ \bibnamefont
  {Landau}},\ }\bibfield  {title} {\enquote {\bibinfo {title} {{The Theory of a
  Fermi Liquid}},}\ }\href
  {http://jetp.ras.ru/cgi-bin/e/index/e/3/6/p920?a=list} {\bibfield  {journal}
  {\bibinfo  {journal} {Sov. Phys. JETP}\ }\textbf {\bibinfo {volume} {3}},\
  \bibinfo {pages} {920} (\bibinfo {year} {1957}{\natexlab{a}})}\BibitemShut
  {NoStop}%
\bibitem [{\citenamefont {Landau}(1957{\natexlab{b}})}]{landau57b}%
  \BibitemOpen
  \bibfield  {author} {\bibinfo {author} {\bibfnamefont {L.~D.}\ \bibnamefont
  {Landau}},\ }\bibfield  {title} {\enquote {\bibinfo {title} {{Oscillations in
  a Fermi Liquid}},}\ }\href
  {http://jetp.ras.ru/cgi-bin/e/index/e/5/1/p101?a=list} {\bibfield  {journal}
  {\bibinfo  {journal} {Sov. Phys. JETP}\ }\textbf {\bibinfo {volume} {5}},\
  \bibinfo {pages} {101} (\bibinfo {year} {1957}{\natexlab{b}})}\BibitemShut
  {NoStop}%
\bibitem [{\citenamefont {Anderson}\ and\ \citenamefont
  {Miller}(2011)}]{anderson11}%
  \BibitemOpen
  \bibfield  {author} {\bibinfo {author} {\bibfnamefont {R.~H.}\ \bibnamefont
  {Anderson}}\ and\ \bibinfo {author} {\bibfnamefont {M.~D.}\ \bibnamefont
  {Miller}},\ }\bibfield  {title} {\enquote {\bibinfo {title} {{Polarization
  dependence of Landau parameters for normal Fermi liquids in two
  dimensions}},}\ }\href {\doibase 10.1103/PhysRevB.84.024504} {\bibfield
  {journal} {\bibinfo  {journal} {Phys. Rev. B}\ }\textbf {\bibinfo {volume}
  {84}},\ \bibinfo {pages} {024504} (\bibinfo {year} {2011})}\BibitemShut
  {NoStop}%
\bibitem [{\citenamefont {Khoo}\ and\ \citenamefont
  {Villadiego}(2019)}]{khoo19}%
  \BibitemOpen
  \bibfield  {author} {\bibinfo {author} {\bibfnamefont {J.~Y.}\ \bibnamefont
  {Khoo}}\ and\ \bibinfo {author} {\bibfnamefont {I.~S.}\ \bibnamefont
  {Villadiego}},\ }\bibfield  {title} {\enquote {\bibinfo {title} {{Shear sound
  of two-dimensional Fermi liquids}},}\ }\href {\doibase
  10.1103/PhysRevB.99.075434} {\bibfield  {journal} {\bibinfo  {journal} {Phys.
  Rev. B}\ }\textbf {\bibinfo {volume} {99}},\ \bibinfo {pages} {075434}
  (\bibinfo {year} {2019})}\BibitemShut {NoStop}%
\bibitem [{\citenamefont {Klein}\ \emph {et~al.}(2019)\citenamefont {Klein},
  \citenamefont {Maslov}, \citenamefont {Pitaevskii},\ and\ \citenamefont
  {Chubukov}}]{klein19}%
  \BibitemOpen
  \bibfield  {author} {\bibinfo {author} {\bibfnamefont {A.}~\bibnamefont
  {Klein}}, \bibinfo {author} {\bibfnamefont {D.~L.}\ \bibnamefont {Maslov}},
  \bibinfo {author} {\bibfnamefont {L.~P.}\ \bibnamefont {Pitaevskii}}, \ and\
  \bibinfo {author} {\bibfnamefont {A.~V.}\ \bibnamefont {Chubukov}},\
  }\bibfield  {title} {\enquote {\bibinfo {title} {{Collective modes near a
  Pomeranchuk instability in two dimensions}},}\ }\href {\doibase
  10.1103/PhysRevResearch.1.033134} {\bibfield  {journal} {\bibinfo  {journal}
  {Phys. Rev. Research}\ }\textbf {\bibinfo {volume} {1}},\ \bibinfo {pages}
  {033134} (\bibinfo {year} {2019})}\BibitemShut {NoStop}%
\bibitem [{\citenamefont {Torre}\ \emph {et~al.}(2019)\citenamefont {Torre},
  \citenamefont {Vieira~de Castro}, \citenamefont {Van~Duppen}, \citenamefont
  {Barcons~Ruiz}, \citenamefont {Peeters}, \citenamefont {Koppens},\ and\
  \citenamefont {Polini}}]{torre19}%
  \BibitemOpen
  \bibfield  {author} {\bibinfo {author} {\bibfnamefont {I.}~\bibnamefont
  {Torre}}, \bibinfo {author} {\bibfnamefont {L.}~\bibnamefont {Vieira~de
  Castro}}, \bibinfo {author} {\bibfnamefont {B.}~\bibnamefont {Van~Duppen}},
  \bibinfo {author} {\bibfnamefont {D.}~\bibnamefont {Barcons~Ruiz}}, \bibinfo
  {author} {\bibfnamefont {F.~M.}\ \bibnamefont {Peeters}}, \bibinfo {author}
  {\bibfnamefont {F.~H.~L.}\ \bibnamefont {Koppens}}, \ and\ \bibinfo {author}
  {\bibfnamefont {M.}~\bibnamefont {Polini}},\ }\bibfield  {title} {\enquote
  {\bibinfo {title} {{Acoustic plasmons at the crossover between the
  collisionless and hydrodynamic regimes in two-dimensional electron
  liquids}},}\ }\href {\doibase 10.1103/PhysRevB.99.144307} {\bibfield
  {journal} {\bibinfo  {journal} {Phys. Rev. B}\ }\textbf {\bibinfo {volume}
  {99}},\ \bibinfo {pages} {144307} (\bibinfo {year} {2019})}\BibitemShut
  {NoStop}%
\bibitem [{\citenamefont {Guo}\ \emph {et~al.}(2016)\citenamefont {Guo},
  \citenamefont {Ilseven}, \citenamefont {Falkovich},\ and\ \citenamefont
  {Levitov}}]{guo16}%
  \BibitemOpen
  \bibfield  {author} {\bibinfo {author} {\bibfnamefont {H.}~\bibnamefont
  {Guo}}, \bibinfo {author} {\bibfnamefont {E.}~\bibnamefont {Ilseven}},
  \bibinfo {author} {\bibfnamefont {G.}~\bibnamefont {Falkovich}}, \ and\
  \bibinfo {author} {\bibfnamefont {L.}~\bibnamefont {Levitov}},\ }\bibfield
  {title} {\enquote {\bibinfo {title} {{Stokes Paradox, Back Reflections and
  Interaction-Enhanced Conduction}},}\ }\href@noop {} {\bibfield  {journal}
  {\bibinfo  {journal} {arXiv:1612.09239}\ } (\bibinfo {year}
  {2016})}\BibitemShut {NoStop}%
\bibitem [{\citenamefont {Lucas}\ and\ \citenamefont
  {Das~Sarma}(2018)}]{lucas18}%
  \BibitemOpen
  \bibfield  {author} {\bibinfo {author} {\bibfnamefont {A.}~\bibnamefont
  {Lucas}}\ and\ \bibinfo {author} {\bibfnamefont {S.}~\bibnamefont
  {Das~Sarma}},\ }\bibfield  {title} {\enquote {\bibinfo {title} {{Electronic
  sound modes and plasmons in hydrodynamic two-dimensional metals}},}\ }\href
  {\doibase 10.1103/PhysRevB.97.115449} {\bibfield  {journal} {\bibinfo
  {journal} {Phys. Rev. B}\ }\textbf {\bibinfo {volume} {97}},\ \bibinfo
  {pages} {115449} (\bibinfo {year} {2018})}\BibitemShut {NoStop}%
\bibitem [{\citenamefont {Abel}\ \emph {et~al.}(1966)\citenamefont {Abel},
  \citenamefont {Anderson},\ and\ \citenamefont {Wheatley}}]{abel66}%
  \BibitemOpen
  \bibfield  {author} {\bibinfo {author} {\bibfnamefont {W.~R.}\ \bibnamefont
  {Abel}}, \bibinfo {author} {\bibfnamefont {A.~C.}\ \bibnamefont {Anderson}},
  \ and\ \bibinfo {author} {\bibfnamefont {J.~C.}\ \bibnamefont {Wheatley}},\
  }\bibfield  {title} {\enquote {\bibinfo {title} {{Propagation of Zero Sound
  in Liquid ${\mathrm{He}}^{3}$ at Low Temperatures}},}\ }\href {\doibase
  10.1103/PhysRevLett.17.74} {\bibfield  {journal} {\bibinfo  {journal} {Phys.
  Rev. Lett.}\ }\textbf {\bibinfo {volume} {17}},\ \bibinfo {pages} {74}
  (\bibinfo {year} {1966})}\BibitemShut {NoStop}%
\bibitem [{\citenamefont {Fr\"ohlich}\ \emph {et~al.}(2012)\citenamefont
  {Fr\"ohlich}, \citenamefont {Feld}, \citenamefont {Vogt}, \citenamefont
  {Koschorreck}, \citenamefont {K\"ohl}, \citenamefont {Berthod},\ and\
  \citenamefont {Giamarchi}}]{froehlich12}%
  \BibitemOpen
  \bibfield  {author} {\bibinfo {author} {\bibfnamefont {B.}~\bibnamefont
  {Fr\"ohlich}}, \bibinfo {author} {\bibfnamefont {M.}~\bibnamefont {Feld}},
  \bibinfo {author} {\bibfnamefont {E.}~\bibnamefont {Vogt}}, \bibinfo {author}
  {\bibfnamefont {M.}~\bibnamefont {Koschorreck}}, \bibinfo {author}
  {\bibfnamefont {M.}~\bibnamefont {K\"ohl}}, \bibinfo {author} {\bibfnamefont
  {C.}~\bibnamefont {Berthod}}, \ and\ \bibinfo {author} {\bibfnamefont
  {T.}~\bibnamefont {Giamarchi}},\ }\bibfield  {title} {\enquote {\bibinfo
  {title} {{Two-Dimensional Fermi Liquid with Attractive Interactions}},}\
  }\href {\doibase 10.1103/PhysRevLett.109.130403} {\bibfield  {journal}
  {\bibinfo  {journal} {Phys. Rev. Lett.}\ }\textbf {\bibinfo {volume} {109}},\
  \bibinfo {pages} {130403} (\bibinfo {year} {2012})}\BibitemShut {NoStop}%
\bibitem [{\citenamefont {Yan}\ \emph {et~al.}(2019)\citenamefont {Yan},
  \citenamefont {Patel}, \citenamefont {Mukherjee}, \citenamefont {Fletcher},
  \citenamefont {Struck},\ and\ \citenamefont {Zwierlein}}]{yan19}%
  \BibitemOpen
  \bibfield  {author} {\bibinfo {author} {\bibfnamefont {Z.}~\bibnamefont
  {Yan}}, \bibinfo {author} {\bibfnamefont {P.~B.}\ \bibnamefont {Patel}},
  \bibinfo {author} {\bibfnamefont {B.}~\bibnamefont {Mukherjee}}, \bibinfo
  {author} {\bibfnamefont {R.~J.}\ \bibnamefont {Fletcher}}, \bibinfo {author}
  {\bibfnamefont {J.}~\bibnamefont {Struck}}, \ and\ \bibinfo {author}
  {\bibfnamefont {M.~W.}\ \bibnamefont {Zwierlein}},\ }\bibfield  {title}
  {\enquote {\bibinfo {title} {{Boiling a Unitary Fermi Liquid}},}\ }\href
  {\doibase 10.1103/PhysRevLett.122.093401} {\bibfield  {journal} {\bibinfo
  {journal} {Phys. Rev. Lett.}\ }\textbf {\bibinfo {volume} {122}},\ \bibinfo
  {pages} {093401} (\bibinfo {year} {2019})}\BibitemShut {NoStop}%
\bibitem [{\citenamefont {Ando}\ \emph {et~al.}(1982)\citenamefont {Ando},
  \citenamefont {Fowler},\ and\ \citenamefont {Stern}}]{ando82}%
  \BibitemOpen
  \bibfield  {author} {\bibinfo {author} {\bibfnamefont {T.}~\bibnamefont
  {Ando}}, \bibinfo {author} {\bibfnamefont {A.~B.}\ \bibnamefont {Fowler}}, \
  and\ \bibinfo {author} {\bibfnamefont {F.}~\bibnamefont {Stern}},\ }\bibfield
   {title} {\enquote {\bibinfo {title} {{Electronic properties of
  two-dimensional systems}},}\ }\href {\doibase 10.1103/RevModPhys.54.437}
  {\bibfield  {journal} {\bibinfo  {journal} {Rev. Mod. Phys.}\ }\textbf
  {\bibinfo {volume} {54}},\ \bibinfo {pages} {437} (\bibinfo {year}
  {1982})}\BibitemShut {NoStop}%
\bibitem [{\citenamefont {Hwang}\ and\ \citenamefont
  {Das~Sarma}(2007)}]{hwang07}%
  \BibitemOpen
  \bibfield  {author} {\bibinfo {author} {\bibfnamefont {E.~H.}\ \bibnamefont
  {Hwang}}\ and\ \bibinfo {author} {\bibfnamefont {S.}~\bibnamefont
  {Das~Sarma}},\ }\bibfield  {title} {\enquote {\bibinfo {title} {{Dielectric
  function, screening, and plasmons in two-dimensional graphene}},}\ }\href
  {\doibase 10.1103/PhysRevB.75.205418} {\bibfield  {journal} {\bibinfo
  {journal} {Phys. Rev. B}\ }\textbf {\bibinfo {volume} {75}},\ \bibinfo
  {pages} {205418} (\bibinfo {year} {2007})}\BibitemShut {NoStop}%
\bibitem [{\citenamefont {Das~Sarma}\ and\ \citenamefont
  {Hwang}(2009)}]{dassarma09}%
  \BibitemOpen
  \bibfield  {author} {\bibinfo {author} {\bibfnamefont {S.}~\bibnamefont
  {Das~Sarma}}\ and\ \bibinfo {author} {\bibfnamefont {E.~H.}\ \bibnamefont
  {Hwang}},\ }\bibfield  {title} {\enquote {\bibinfo {title} {{Collective Modes
  of the Massless Dirac Plasma}},}\ }\href {\doibase
  10.1103/PhysRevLett.102.206412} {\bibfield  {journal} {\bibinfo  {journal}
  {Phys. Rev. Lett.}\ }\textbf {\bibinfo {volume} {102}},\ \bibinfo {pages}
  {206412} (\bibinfo {year} {2009})}\BibitemShut {NoStop}%
\bibitem [{\citenamefont {Conti}\ and\ \citenamefont
  {Vignale}(1999)}]{conti99}%
  \BibitemOpen
  \bibfield  {author} {\bibinfo {author} {\bibfnamefont {S.}~\bibnamefont
  {Conti}}\ and\ \bibinfo {author} {\bibfnamefont {G.}~\bibnamefont
  {Vignale}},\ }\bibfield  {title} {\enquote {\bibinfo {title} {{Elasticity of
  an electron liquid}},}\ }\href {\doibase 10.1103/PhysRevB.60.7966} {\bibfield
   {journal} {\bibinfo  {journal} {Phys. Rev. B}\ }\textbf {\bibinfo {volume}
  {60}},\ \bibinfo {pages} {7966} (\bibinfo {year} {1999})}\BibitemShut
  {NoStop}%
\bibitem [{\citenamefont {Gao}\ \emph {et~al.}(2010)\citenamefont {Gao},
  \citenamefont {Tao}, \citenamefont {Vignale},\ and\ \citenamefont
  {Tokatly}}]{gao10}%
  \BibitemOpen
  \bibfield  {author} {\bibinfo {author} {\bibfnamefont {X.}~\bibnamefont
  {Gao}}, \bibinfo {author} {\bibfnamefont {J.}~\bibnamefont {Tao}}, \bibinfo
  {author} {\bibfnamefont {G.}~\bibnamefont {Vignale}}, \ and\ \bibinfo
  {author} {\bibfnamefont {I.~V.}\ \bibnamefont {Tokatly}},\ }\bibfield
  {title} {\enquote {\bibinfo {title} {{Continuum mechanics for quantum
  many-body systems: Linear response regime}},}\ }\href {\doibase
  10.1103/PhysRevB.81.195106} {\bibfield  {journal} {\bibinfo  {journal} {Phys.
  Rev. B}\ }\textbf {\bibinfo {volume} {81}},\ \bibinfo {pages} {195106}
  (\bibinfo {year} {2010})}\BibitemShut {NoStop}%
\bibitem [{\citenamefont {Klein}\ \emph {et~al.}(2020)\citenamefont {Klein},
  \citenamefont {Maslov},\ and\ \citenamefont {Chubukov}}]{klein20}%
  \BibitemOpen
  \bibfield  {author} {\bibinfo {author} {\bibfnamefont {A.}~\bibnamefont
  {Klein}}, \bibinfo {author} {\bibfnamefont {D.~L.}\ \bibnamefont {Maslov}}, \
  and\ \bibinfo {author} {\bibfnamefont {A.~V.}\ \bibnamefont {Chubukov}},\
  }\bibfield  {title} {\enquote {\bibinfo {title} {{Hidden and mirage
  collective modes in two dimensional Fermi liquids}},}\ }\href {\doibase
  10.1038/s41535-020-0250-4} {\bibfield  {journal} {\bibinfo  {journal} {npj
  Quantum Materials}\ }\textbf {\bibinfo {volume} {5}},\ \bibinfo {pages} {55}
  (\bibinfo {year} {2020})}\BibitemShut {NoStop}%
\bibitem [{\citenamefont {Vogt}\ \emph {et~al.}(2012)\citenamefont {Vogt},
  \citenamefont {Feld}, \citenamefont {Fr\"ohlich}, \citenamefont {Pertot},
  \citenamefont {Koschorreck},\ and\ \citenamefont {K\"ohl}}]{vogt12}%
  \BibitemOpen
  \bibfield  {author} {\bibinfo {author} {\bibfnamefont {E.}~\bibnamefont
  {Vogt}}, \bibinfo {author} {\bibfnamefont {M.}~\bibnamefont {Feld}}, \bibinfo
  {author} {\bibfnamefont {B.}~\bibnamefont {Fr\"ohlich}}, \bibinfo {author}
  {\bibfnamefont {D.}~\bibnamefont {Pertot}}, \bibinfo {author} {\bibfnamefont
  {M.}~\bibnamefont {Koschorreck}}, \ and\ \bibinfo {author} {\bibfnamefont
  {M.}~\bibnamefont {K\"ohl}},\ }\bibfield  {title} {\enquote {\bibinfo {title}
  {{Scale Invariance and Viscosity of a Two-Dimensional Fermi Gas}},}\ }\href
  {\doibase 10.1103/PhysRevLett.108.070404} {\bibfield  {journal} {\bibinfo
  {journal} {Phys. Rev. Lett.}\ }\textbf {\bibinfo {volume} {108}},\ \bibinfo
  {pages} {070404} (\bibinfo {year} {2012})}\BibitemShut {NoStop}%
\bibitem [{\citenamefont {Patel}\ \emph {et~al.}(2020)\citenamefont {Patel},
  \citenamefont {Yan}, \citenamefont {Mukherjee}, \citenamefont {Fletcher},
  \citenamefont {Struck},\ and\ \citenamefont {Zwierlein}}]{patel20}%
  \BibitemOpen
  \bibfield  {author} {\bibinfo {author} {\bibfnamefont {P.~B.}\ \bibnamefont
  {Patel}}, \bibinfo {author} {\bibfnamefont {Z.}~\bibnamefont {Yan}}, \bibinfo
  {author} {\bibfnamefont {B.}~\bibnamefont {Mukherjee}}, \bibinfo {author}
  {\bibfnamefont {R.~J.}\ \bibnamefont {Fletcher}}, \bibinfo {author}
  {\bibfnamefont {J.}~\bibnamefont {Struck}}, \ and\ \bibinfo {author}
  {\bibfnamefont {M.~W.}\ \bibnamefont {Zwierlein}},\ }\bibfield  {title}
  {\enquote {\bibinfo {title} {{Universal sound diffusion in a strongly
  interacting Fermi gas}},}\ }\href {\doibase 10.1126/science.aaz5756}
  {\bibfield  {journal} {\bibinfo  {journal} {Science}\ }\textbf {\bibinfo
  {volume} {370}},\ \bibinfo {pages} {1222} (\bibinfo {year}
  {2020})}\BibitemShut {NoStop}%
\bibitem [{\citenamefont {Allen}\ \emph {et~al.}(1977)\citenamefont {Allen},
  \citenamefont {Tsui},\ and\ \citenamefont {Logan}}]{allen77}%
  \BibitemOpen
  \bibfield  {author} {\bibinfo {author} {\bibfnamefont {S.~J.}\ \bibnamefont
  {Allen}}, \bibinfo {author} {\bibfnamefont {D.~C.}\ \bibnamefont {Tsui}}, \
  and\ \bibinfo {author} {\bibfnamefont {R.~A.}\ \bibnamefont {Logan}},\
  }\bibfield  {title} {\enquote {\bibinfo {title} {{Observation of the
  Two-Dimensional Plasmon in Silicon Inversion Layers}},}\ }\href {\doibase
  10.1103/PhysRevLett.38.980} {\bibfield  {journal} {\bibinfo  {journal} {Phys.
  Rev. Lett.}\ }\textbf {\bibinfo {volume} {38}},\ \bibinfo {pages} {980}
  (\bibinfo {year} {1977})}\BibitemShut {NoStop}%
\bibitem [{\citenamefont {Pinczuk}\ \emph {et~al.}(1981)\citenamefont
  {Pinczuk}, \citenamefont {Shah},\ and\ \citenamefont {Wolff}}]{pinczuk81}%
  \BibitemOpen
  \bibfield  {author} {\bibinfo {author} {\bibfnamefont {A.}~\bibnamefont
  {Pinczuk}}, \bibinfo {author} {\bibfnamefont {J.}~\bibnamefont {Shah}}, \
  and\ \bibinfo {author} {\bibfnamefont {P.~A.}\ \bibnamefont {Wolff}},\
  }\bibfield  {title} {\enquote {\bibinfo {title} {{Collective Modes of
  Photoexcited Electron-Hole Plasmas in GaAs}},}\ }\href {\doibase
  10.1103/PhysRevLett.47.1487} {\bibfield  {journal} {\bibinfo  {journal}
  {Phys. Rev. Lett.}\ }\textbf {\bibinfo {volume} {47}},\ \bibinfo {pages}
  {1487} (\bibinfo {year} {1981})}\BibitemShut {NoStop}%
\bibitem [{\citenamefont {Pinczuk}\ \emph {et~al.}(1986)\citenamefont
  {Pinczuk}, \citenamefont {Lamont},\ and\ \citenamefont
  {Gossard}}]{pinczuk86}%
  \BibitemOpen
  \bibfield  {author} {\bibinfo {author} {\bibfnamefont {A.}~\bibnamefont
  {Pinczuk}}, \bibinfo {author} {\bibfnamefont {M.~G.}\ \bibnamefont {Lamont}},
  \ and\ \bibinfo {author} {\bibfnamefont {A.~C.}\ \bibnamefont {Gossard}},\
  }\bibfield  {title} {\enquote {\bibinfo {title} {{Discrete Plasmons in Finite
  Semiconductor Multilayers}},}\ }\href {\doibase 10.1103/PhysRevLett.56.2092}
  {\bibfield  {journal} {\bibinfo  {journal} {Phys. Rev. Lett.}\ }\textbf
  {\bibinfo {volume} {56}},\ \bibinfo {pages} {2092} (\bibinfo {year}
  {1986})}\BibitemShut {NoStop}%
\bibitem [{\citenamefont {Bostwick}\ \emph {et~al.}(2007)\citenamefont
  {Bostwick}, \citenamefont {Ohta}, \citenamefont {Seyller}, \citenamefont
  {Horn},\ and\ \citenamefont {Rotenberg}}]{bostwick07}%
  \BibitemOpen
  \bibfield  {author} {\bibinfo {author} {\bibfnamefont {A.}~\bibnamefont
  {Bostwick}}, \bibinfo {author} {\bibfnamefont {T.}~\bibnamefont {Ohta}},
  \bibinfo {author} {\bibfnamefont {T.}~\bibnamefont {Seyller}}, \bibinfo
  {author} {\bibfnamefont {K.}~\bibnamefont {Horn}}, \ and\ \bibinfo {author}
  {\bibfnamefont {E.}~\bibnamefont {Rotenberg}},\ }\bibfield  {title} {\enquote
  {\bibinfo {title} {{Quasiparticle dynamics in graphene}},}\ }\href {\doibase
  10.1038/nphys477} {\bibfield  {journal} {\bibinfo  {journal} {Nature
  Physics}\ }\textbf {\bibinfo {volume} {3}},\ \bibinfo {pages} {36} (\bibinfo
  {year} {2007})}\BibitemShut {NoStop}%
\bibitem [{\citenamefont {Liu}\ \emph {et~al.}(2008)\citenamefont {Liu},
  \citenamefont {Willis}, \citenamefont {Emtsev},\ and\ \citenamefont
  {Seyller}}]{liu08}%
  \BibitemOpen
  \bibfield  {author} {\bibinfo {author} {\bibfnamefont {Y.}~\bibnamefont
  {Liu}}, \bibinfo {author} {\bibfnamefont {R.~F.}\ \bibnamefont {Willis}},
  \bibinfo {author} {\bibfnamefont {K.~V.}\ \bibnamefont {Emtsev}}, \ and\
  \bibinfo {author} {\bibfnamefont {Th.}\ \bibnamefont {Seyller}},\ }\bibfield
  {title} {\enquote {\bibinfo {title} {{Plasmon dispersion and damping in
  electrically isolated two-dimensional charge sheets}},}\ }\href {\doibase
  10.1103/PhysRevB.78.201403} {\bibfield  {journal} {\bibinfo  {journal} {Phys.
  Rev. B}\ }\textbf {\bibinfo {volume} {78}},\ \bibinfo {pages} {201403}
  (\bibinfo {year} {2008})}\BibitemShut {NoStop}%
\bibitem [{\citenamefont {Tegenkamp}\ \emph {et~al.}(2010)\citenamefont
  {Tegenkamp}, \citenamefont {Pfn\"{u}r}, \citenamefont {Langer}, \citenamefont
  {Baringhaus},\ and\ \citenamefont {Schumacher}}]{tegenkamp10}%
  \BibitemOpen
  \bibfield  {author} {\bibinfo {author} {\bibfnamefont {C.}~\bibnamefont
  {Tegenkamp}}, \bibinfo {author} {\bibfnamefont {H.}~\bibnamefont
  {Pfn\"{u}r}}, \bibinfo {author} {\bibfnamefont {T.}~\bibnamefont {Langer}},
  \bibinfo {author} {\bibfnamefont {J.}~\bibnamefont {Baringhaus}}, \ and\
  \bibinfo {author} {\bibfnamefont {H.~W.}\ \bibnamefont {Schumacher}},\
  }\bibfield  {title} {\enquote {\bibinfo {title} {{Plasmon
  electron{\textendash}hole resonance in epitaxial graphene}},}\ }\href
  {\doibase 10.1088/0953-8984/23/1/012001} {\bibfield  {journal} {\bibinfo
  {journal} {Journal of Physics: Condensed Matter}\ }\textbf {\bibinfo {volume}
  {23}},\ \bibinfo {pages} {012001} (\bibinfo {year} {2010})}\BibitemShut
  {NoStop}%
\bibitem [{\citenamefont {Bostwick}\ \emph {et~al.}(2010)\citenamefont
  {Bostwick}, \citenamefont {Speck}, \citenamefont {Seyller}, \citenamefont
  {Horn}, \citenamefont {Polini}, \citenamefont {Asgari}, \citenamefont
  {MacDonald},\ and\ \citenamefont {Rotenberg}}]{bostwick10}%
  \BibitemOpen
  \bibfield  {author} {\bibinfo {author} {\bibfnamefont {A.}~\bibnamefont
  {Bostwick}}, \bibinfo {author} {\bibfnamefont {F.}~\bibnamefont {Speck}},
  \bibinfo {author} {\bibfnamefont {T.}~\bibnamefont {Seyller}}, \bibinfo
  {author} {\bibfnamefont {K.}~\bibnamefont {Horn}}, \bibinfo {author}
  {\bibfnamefont {M.}~\bibnamefont {Polini}}, \bibinfo {author} {\bibfnamefont
  {R.}~\bibnamefont {Asgari}}, \bibinfo {author} {\bibfnamefont {A.~H.}\
  \bibnamefont {MacDonald}}, \ and\ \bibinfo {author} {\bibfnamefont
  {E.}~\bibnamefont {Rotenberg}},\ }\bibfield  {title} {\enquote {\bibinfo
  {title} {{Observation of Plasmarons in Quasi-Freestanding Doped Graphene}},}\
  }\href {\doibase 10.1126/science.1186489} {\bibfield  {journal} {\bibinfo
  {journal} {Science}\ }\textbf {\bibinfo {volume} {328}},\ \bibinfo {pages}
  {999} (\bibinfo {year} {2010})}\BibitemShut {NoStop}%
\bibitem [{\citenamefont {Ju}\ \emph {et~al.}(2011)\citenamefont {Ju},
  \citenamefont {Geng}, \citenamefont {Horng}, \citenamefont {Girit},
  \citenamefont {Martin}, \citenamefont {Hao}, \citenamefont {Bechtel},
  \citenamefont {Liang}, \citenamefont {Zettl}, \citenamefont {Shen},\ and\
  \citenamefont {Wang}}]{ju11}%
  \BibitemOpen
  \bibfield  {author} {\bibinfo {author} {\bibfnamefont {L.}~\bibnamefont
  {Ju}}, \bibinfo {author} {\bibfnamefont {B.}~\bibnamefont {Geng}}, \bibinfo
  {author} {\bibfnamefont {J.}~\bibnamefont {Horng}}, \bibinfo {author}
  {\bibfnamefont {C.}~\bibnamefont {Girit}}, \bibinfo {author} {\bibfnamefont
  {M.}~\bibnamefont {Martin}}, \bibinfo {author} {\bibfnamefont
  {Z.}~\bibnamefont {Hao}}, \bibinfo {author} {\bibfnamefont {H.~A.}\
  \bibnamefont {Bechtel}}, \bibinfo {author} {\bibfnamefont {X.}~\bibnamefont
  {Liang}}, \bibinfo {author} {\bibfnamefont {A.}~\bibnamefont {Zettl}},
  \bibinfo {author} {\bibfnamefont {Y.~R.}\ \bibnamefont {Shen}}, \ and\
  \bibinfo {author} {\bibfnamefont {F.}~\bibnamefont {Wang}},\ }\bibfield
  {title} {\enquote {\bibinfo {title} {{Graphene plasmonics for tunable
  terahertz metamaterials}},}\ }\href {\doibase 10.1038/nnano.2011.146}
  {\bibfield  {journal} {\bibinfo  {journal} {Nature Nanotechnology}\ }\textbf
  {\bibinfo {volume} {6}},\ \bibinfo {pages} {630} (\bibinfo {year}
  {2011})}\BibitemShut {NoStop}%
\bibitem [{\citenamefont {Roach}\ and\ \citenamefont
  {Ketterson}(1976)}]{roach76}%
  \BibitemOpen
  \bibfield  {author} {\bibinfo {author} {\bibfnamefont {P.~R.}\ \bibnamefont
  {Roach}}\ and\ \bibinfo {author} {\bibfnamefont {J.~B.}\ \bibnamefont
  {Ketterson}},\ }\bibfield  {title} {\enquote {\bibinfo {title} {{Observation
  of Transverse Zero Sound in Normal $^{3}\mathrm{He}$}},}\ }\href {\doibase
  10.1103/PhysRevLett.36.736} {\bibfield  {journal} {\bibinfo  {journal} {Phys.
  Rev. Lett.}\ }\textbf {\bibinfo {volume} {36}},\ \bibinfo {pages} {736}
  (\bibinfo {year} {1976})}\BibitemShut {NoStop}%
\bibitem [{\citenamefont {Flowers}\ \emph {et~al.}(1976)\citenamefont
  {Flowers}, \citenamefont {Richardson},\ and\ \citenamefont
  {Williamson}}]{flowers76}%
  \BibitemOpen
  \bibfield  {author} {\bibinfo {author} {\bibfnamefont {E.~G.}\ \bibnamefont
  {Flowers}}, \bibinfo {author} {\bibfnamefont {R.~W.}\ \bibnamefont
  {Richardson}}, \ and\ \bibinfo {author} {\bibfnamefont {S.~J.}\ \bibnamefont
  {Williamson}},\ }\bibfield  {title} {\enquote {\bibinfo {title} {{Transverse
  Zero Sound in Normal $^{3}\mathrm{He}$}},}\ }\href {\doibase
  10.1103/PhysRevLett.37.309} {\bibfield  {journal} {\bibinfo  {journal} {Phys.
  Rev. Lett.}\ }\textbf {\bibinfo {volume} {37}},\ \bibinfo {pages} {309}
  (\bibinfo {year} {1976})}\BibitemShut {NoStop}%
\bibitem [{\citenamefont {Khoo}\ \emph {et~al.}(2020)\citenamefont {Khoo},
  \citenamefont {Chang}, \citenamefont {Pientka},\ and\ \citenamefont
  {Sodemann}}]{khoo20}%
  \BibitemOpen
  \bibfield  {author} {\bibinfo {author} {\bibfnamefont {J.~Y.}\ \bibnamefont
  {Khoo}}, \bibinfo {author} {\bibfnamefont {P.-Y.}\ \bibnamefont {Chang}},
  \bibinfo {author} {\bibfnamefont {F.}~\bibnamefont {Pientka}}, \ and\
  \bibinfo {author} {\bibfnamefont {I.}~\bibnamefont {Sodemann}},\ }\bibfield
  {title} {\enquote {\bibinfo {title} {{Quantum paracrystalline shear modes of
  the electron liquid}},}\ }\href {\doibase 10.1103/PhysRevB.102.085437}
  {\bibfield  {journal} {\bibinfo  {journal} {Phys. Rev. B}\ }\textbf {\bibinfo
  {volume} {102}},\ \bibinfo {pages} {085437} (\bibinfo {year}
  {2020})}\BibitemShut {NoStop}%
\bibitem [{\citenamefont {Valentinis}\ \emph {et~al.}(2021)\citenamefont
  {Valentinis}, \citenamefont {Zaanen},\ and\ \citenamefont {van~der
  Marel}}]{valentinis21}%
  \BibitemOpen
  \bibfield  {author} {\bibinfo {author} {\bibfnamefont {D.}~\bibnamefont
  {Valentinis}}, \bibinfo {author} {\bibfnamefont {J.}~\bibnamefont {Zaanen}},
  \ and\ \bibinfo {author} {\bibfnamefont {D.}~\bibnamefont {van~der Marel}},\
  }\bibfield  {title} {\enquote {\bibinfo {title} {{Propagation of shear stress
  in strongly interacting metallic Fermi liquids enhances transmission of
  terahertz radiation}},}\ }\href {\doibase 10.1038/s41598-021-86356-2}
  {\bibfield  {journal} {\bibinfo  {journal} {Scientific Reports}\ }\textbf
  {\bibinfo {volume} {11}},\ \bibinfo {pages} {7105} (\bibinfo {year}
  {2021})}\BibitemShut {NoStop}%
\bibitem [{\citenamefont {Kryhin}\ and\ \citenamefont
  {Levitov}(2022)}]{kryhin22}%
  \BibitemOpen
  \bibfield  {author} {\bibinfo {author} {\bibfnamefont {S.}~\bibnamefont
  {Kryhin}}\ and\ \bibinfo {author} {\bibfnamefont {L.}~\bibnamefont
  {Levitov}},\ }\bibfield  {title} {\enquote {\bibinfo {title} {{Collinear
  scattering and long-time dynamical memory in two-dimensional electron
  fluids}},}\ }\href@noop {} {\bibfield  {journal} {\bibinfo  {journal}
  {arXiv:2112.05076v3}\ } (\bibinfo {year} {2022})}\BibitemShut {NoStop}%
\bibitem [{\citenamefont {Setiawan}\ and\ \citenamefont
  {Hofmann}(2022)}]{setiawan21}%
  \BibitemOpen
  \bibfield  {author} {\bibinfo {author} {\bibfnamefont {F.}~\bibnamefont
  {Setiawan}}\ and\ \bibinfo {author} {\bibfnamefont {J.}~\bibnamefont
  {Hofmann}},\ }\bibfield  {title} {\enquote {\bibinfo {title} {{Analytic
  approach to transport in superconducting junctions with arbitrary carrier
  density}},}\ }\href {\doibase 10.1103/PhysRevResearch.4.043087} {\bibfield
  {journal} {\bibinfo  {journal} {Phys. Rev. Research}\ }\textbf {\bibinfo
  {volume} {4}},\ \bibinfo {pages} {043087} (\bibinfo {year}
  {2022})}\BibitemShut {NoStop}%
\bibitem [{\citenamefont {Press}\ \emph {et~al.}(2017)\citenamefont {Press},
  \citenamefont {Teukolsky}, \citenamefont {Vetterling},\ and\ \citenamefont
  {Flannery}}]{press17}%
  \BibitemOpen
  \bibfield  {author} {\bibinfo {author} {\bibfnamefont {W.~H.}\ \bibnamefont
  {Press}}, \bibinfo {author} {\bibfnamefont {S.~A.}\ \bibnamefont
  {Teukolsky}}, \bibinfo {author} {\bibfnamefont {W.~T.}\ \bibnamefont
  {Vetterling}}, \ and\ \bibinfo {author} {\bibfnamefont {B.~P.}\ \bibnamefont
  {Flannery}},\ }\href@noop {} {\emph {\bibinfo {title} {{Numerical Recipes:
  The Art of Scientific Computing}}}},\ \bibinfo {edition} {{Third Edition}}\
  ed.\ (\bibinfo  {publisher} {Cambridge University Press, Cambridge},\
  \bibinfo {year} {2017})\BibitemShut {NoStop}%
\end{thebibliography}%

\end{document}